\documentclass[prd,aps,nofootinbib,showpacs]{revtex4}
\usepackage{amsmath,graphicx,color,epsfig}

\begin{document}
\title{ \hfill{\small DESY 10-234}\\[0.5em] 
Branching ratios, forward-backward asymmetries and angular distributions of $B\to K_2^*l^+l^-$ in the standard model and two new physics scenarios}
\author{ Run-Hui Li$^{a,b}$, Cai-Dian L\"u $^a$ and Wei Wang $^c$\footnote{
Alexander von Humboldt Fellow}~\footnote{Email:wei.wang@desy.de}}

\affiliation{
 \it  $^a$ Institute of High Energy Physics, P.O. Box 918(4), Beijing 100049, People's Republic of China\\
 \it  $^b$ Department of Physics $\&$ IPAP, Yonsei University, Seoul 120-479, Korea \\
 \it  $^c$ Deutsches Elektronen-Synchrotron DESY, Hamburg 22607, Germany}

\begin{abstract}
We analyze the  $B\to K_2^*(\to K\pi)l^+l^-$ (with $l=e,\mu,\tau$)
decay in the standard model and two new physics scenarios:
vector-like quark model and family non-universal $Z'$ model. We
derive its differential angular distributions, using the recently calculated form factors in the
perturbative QCD approach. Branching ratios, polarizations,
forward-backward asymmetries and transversity amplitudes are
predicted, from which we find a promising  prospective to observe this
channel in the future experiment.  We   update
the constraints on effective Wilson coefficients and/or free
parameters in these two new physics scenarios by making use of the $B\to K^*l^+l^-$ and $b\to sl^+l^-$ experimental data. Their
impact on $B\to K_2^*l^+l^-$ is subsequently explored and in particular the
zero-crossing point for the forward-backward asymmetry in these new
physics scenarios can sizably deviate from the standard model.  In addition we
also generalize the analysis to a similar mode $B_s\to
f_2'(1525)(\to K^+K^-) l^+l^-$.
\end{abstract}
\maketitle

\section{Introduction}\label{section:introduction}

Discoveries of new degrees of freedom at TeV energy scale, with
contributions to our understanding  of the origin of the electroweak
symmetry breaking, can proceed in two different ways. One is a
direct search of the Higgs boson, the last piece to complete
the standard model (SM), and particles beyond the SM, to
establish new physics (NP) theories. The other effort, already
ongoing, is to investigate processes in which SM is tested with
higher experimental and theoretical precision. Among the latter
category, rare $B$ decays are among ideal probes. Besides
constraints on the Cabibbo-Kobayashi-Maskawa (CKM) matrix including
apexs and angles of the unitary triangle, which have been
contributed by semileptonic $b\to u/c$ and nonleptonic $B$
decays respectively, the electroweak interaction structure can also
be probed by, for instance, the $b\to s\gamma$ and $b\to
sl^+l^-$ modes which are induced by loop effects in the SM and
therefore   sensitive to the NP interactions.

Unlike $b\to s\gamma$ and $B\to K^*\gamma$ that has only limited
physical observables, $b\to sl^+l^-$ especially $B\to
K^*l^+l^-$, with a number of observables accessible, provides a
wealth of information of weak interactions,  ranging from the
forward-backward asymmetries (FBAs), isospin symmetries,
polarizations to a full angular analysis. The last barrier to access this
mode, the low statistic with a branching faction of the order
$10^{-6}$, is being cleared by the  $B$ factories and the
hadron collider~\cite{Aubert:2008ps,:2009zv,Aaltonen:2008xf}. The ongoing
LHCb experiment can accumulate $6200$ events per nominal running
year of $2 fb^{-1}$ with $\sqrt s=14$ TeV~\cite{:2009ny}, which
allows to probe the short-distance physics at an unprecedented
level. For instance the sensitivity to zero-crossing point of FBAs
can be reduced to $0.5{\rm GeV}^2$ and might be further improved as
$0.1{\rm GeV}^2$ after the upgrade~\cite{Gac:2010ws}. This provides
a good sensitivity to discriminate between the SM and different
models of new physics. There are also a lot of opportunities on the
Super B factory~\cite{O'Leary:2010af}. Because of these virtues,
theoretical research interests in this mode have exploded and the
precision is highly improved, see
Refs.~\cite{Ali:1999mm,Kim:2000dq,Beneke:2001at,Chen:2002bq,
Kruger:2005ep,Ali:2006ew,Bobeth:2008ij,Egede:2008uy,Altmannshofer:2008dz,Chiang:2009dx,Alok:2009tz,Chang:2010zy,Bharucha:2010bb,Khodjamirian:2010vf,Bobeth:2010wg,Alok:2010zd}
for an incomplete list.

Toward the direction to elucidate the electroweak interaction, $B\to K^*l^+l^-$ and its SU(3)-related mode
$B_s\to \phi l^+l^-$ are not unique. In this work, we shall
point out that $B\to K_2^*(1430)l^+l^-$ and the $B_s$-counterpart
$B_s\to f_2'(1525)l^+l^-$~\footnote{Hereafter will use $K_2^*$ and
$f_2'$ to abbreviate $K_2^*(1430)$ and $f_2'(1525)$. }, which so far
have not been investigated in
detail~\cite{Rai-Choudhury:2006gv,Hatanaka:2009gb,Hatanaka:2009sj,Choudhury:2009fz},
are also useful in several aspects. Due to the similarities between $K^*$
and $K_2^*$, all experiment techniques for $B\to K^*l^+l^-$ are
adjustable to $B\to K_2^*l^+l^-$.  The main decay
product of $K_2^*$ is a pair of charged kaon and pion which are
easily detected on the LHCb. Moreover as we will show in the
following, based on either a direct computation in the perturbative QCD approach~\cite{Wang:2010ni} or the implication of experimental data on $B\to
K_2^*\gamma$ process, the branching ratio (BR) of $B\to K_2^*
l^+l^-$ is found sizable. Therefore thousands of signal events can
be accumulated on the LHCb per nominal running year.

As a consequence of the unitarity of quark mixing matrix, tree
level flavor-changing neutral-current (FCNC) is forbidden in the SM.
When higher order corrections are taken into account, $b\to sl^+l^-$
arises from photonic penguin, Z penguin and W-box diagram. The large
mass scale of virtual states leads to tiny Wilson coefficients
in $b$ quark decays and thus $b\to sl^+l^-$ would be sensitive to
the potential NP effects. In certain NP scenarios, new effective
operators out of the SM scope can emerge, but in a class of
other scenarios, only Wilson coefficients for effective
operators are modified. Among the latter category, vector-like quark
model
(VQM)~\cite{Rizzo:1985db,Shin:1988eu,Branco:1986my,Langacker:1988ur,Nir:1990yq,Barenboim:1997pf,Barenboim:2001fd,Chen:2008ug,Mohanta:2008fa}
and family non-universal $Z'$
model~\cite{Langacker:2000ju,Barger:2003hg,Chiang:2006we,Barger:2009eq,ZprPheno3,ZprReview}
are simplest and therefore of theoretical interest. In this work we shall
also elaborate the impacts of these models on $B\to K_2^*l^+l^-$.

The rest of the paper is organized as follows. In
Sec.~\ref{section:elements}, we collect the necessary hadronic inputs, namely form factors.
Sec.~II contains the analytic formulas for differential decay
distributions and integrated quantities. In
Sec.~\ref{sec:two-NP-models}, we give a brief overview of two NP
models whose effects we will study. Sec.~V is our
phenomenological analysis: the predictions in the SM; update of the
constraints on the VQM and $Z'$ model parameters; the NP effect on
the physical quantities. We conclude in the last section. In the appendix, we give the effective Hamiltonian 
in the SM and the helicity amplitude method.

\section{$B\to K_2$ form factors}\label{section:elements}
%

$B\to K_2^*l^+l^-$ decay amplitudes contain two separate parts. 
Short-distance physics, in which contributions  at the weak
scale $\mu_W$ is calculated by perturbation theory and the
evolution between $m_W$ and $b$ quark mass scale $m_b$ is organized
by the renormalization group. These degrees of freedom are
incorporated into Wilson coefficients and  the obtained effective Hamiltonian responsible for $b\to sl^+l^-$ in the appendix A.  
The low-energy effect characterizes the 
long-distance physics and  will be
parameterized by hadronic matrix elements of effective
operators, which  are usually reduced to heavy-to-light form factors
in semileptonic $B$ decays.


The spin-2 polarization tensor, which satisfies $\epsilon_{\mu\nu}
P^{\nu}_2=0$ with $P_2$ being the momentum, is symmetric and
traceless. It can be constructed via the spin-1 polarization vector
$\epsilon$:
\begin{eqnarray}
 &&\epsilon_{\mu\nu}(\pm2)=
 \epsilon_\mu(\pm)\epsilon_\nu(\pm),\;\;\;\;
 \epsilon_{\mu\nu}(\pm1)=\frac{1}{\sqrt2}
 [\epsilon_{\mu}(\pm)\epsilon_\nu(0)+\epsilon_{\nu}(\pm)\epsilon_\mu(0)],\nonumber\\
 &&\epsilon_{\mu\nu}(0)=\frac{1}{\sqrt6}
 [\epsilon_{\mu}(+)\epsilon_\nu(-)+\epsilon_{\nu}(+)\epsilon_\mu(-)]
 +\sqrt{\frac{2}{3}}\epsilon_{\mu}(0)\epsilon_\nu(0).
\end{eqnarray}
In the case of the tensor meson moving on the $z$ axis, the explicit
structures of $\epsilon$ in the ordinary coordinate frame are chosen
as
\begin{eqnarray}
\epsilon_\mu(0)&=&\frac{1}{m_{K_2^*}}(|\vec
p_{K_2^*}|,0,0,E_{K_2^*}),\;\;\;
\epsilon_\mu(\pm)=\frac{1}{\sqrt{2}}(0,\mp1,-i,0),
\end{eqnarray}
where $E_{K_2^*}$ and $\vec{p}_{K_2^*}$ is the energy and the
momentum magnitude of ${K_2^*}$ in  $B$ meson rest frame, respectively.
In the following calculation, it is convenient to introduce a new
polarization vector $\epsilon_T$
\begin{eqnarray}
  &&\epsilon_{T\mu}(h) =\frac{1}{m_B}
  \epsilon_{\mu\nu}(h)P_{B}^\nu,
\end{eqnarray}
which satisfies
\begin{eqnarray}
  && \epsilon_{T\mu}(\pm2)=0,\;\;\;
  \epsilon_{T\mu}(\pm1)=\frac{1}{m_B}\frac{1}{\sqrt2}\epsilon(0)\cdot
  P_{B}\epsilon_\mu(\pm),\;\;\;
  \epsilon_{T\mu}(0)=\frac{1}{m_B}\sqrt{\frac{2}{3}}\epsilon(0)\cdot
  P_{B}\epsilon_\mu(0).
\end{eqnarray}
The contraction is evaluated as $\epsilon(0)\cdot P_{B}/m_B=|\vec
p_{K_2^*}|/m_{K_2^*}$ and thus we can see that the new vector
$\epsilon_T$ plays a similar role to the ordinary polarization
vector $\epsilon$, regardless of the dimensionless constants
$\frac{1}{\sqrt2}|\vec p_{K_2^*}|/m_{K_2^*}$ or
$\sqrt{\frac{2}{3}}|\vec p_{K_2^*}|/m_{K_2^*}$.

The parametrization of $B\to K_2^*$ form factors is analogous to
the $B\to K^*$
ones~\cite{Hatanaka:2009gb,Hatanaka:2009sj,Wang:2010ni,Yang:2010qd}
 \begin{eqnarray}
  \langle K_2^*(P_2,\epsilon)|\bar s\gamma^{\mu}b|\overline B(P_B)\rangle
   &=&-\frac{2V(q^2)}{m_B+m_{K_2^*}}\epsilon^{\mu\nu\rho\sigma} \epsilon^*_{T\nu}  P_{B\rho}P_{2\sigma}, \nonumber\\
  \langle  K_2^*(P_2,\epsilon)|\bar s\gamma^{\mu}\gamma_5 b|\overline
  B(P_B)\rangle
   &=&2im_{K_2^*} A_0(q^2)\frac{\epsilon^*_{T } \cdot  q }{ q^2}q^{\mu}
    +i(m_B+m_{K_2^*})A_1(q^2)\left[ \epsilon^*_{T\mu }
    -\frac{\epsilon^*_{T } \cdot  q }{q^2}q^{\mu} \right] \nonumber\\
    &&-iA_2(q^2)\frac{\epsilon^*_{T} \cdot  q }{  m_B+m_{K_2^*} }
     \left[ P^{\mu}-\frac{m_B^2-m_{K_2^*}^2}{q^2}q^{\mu} \right],\nonumber\\
  \langle  K_2^*(P_2,\epsilon)|\bar s\sigma^{\mu\nu}q_{\nu}b|\overline
  B(P_B)\rangle
   &=&-2iT_1(q^2)\epsilon^{\mu\nu\rho\sigma} \epsilon^*_{T\nu} P_{B\rho}P_{2\sigma}, \nonumber\\
  \langle  K_2^*(P_2,\epsilon)|\bar s\sigma^{\mu\nu}\gamma_5q_{\nu}b|\overline  B(P_B)\rangle
   &=&T_2(q^2)\left[(m_B^2-m_{K_2^*}^2) \epsilon^*_{T\mu }
       - {\epsilon^*_{T } \cdot  q }  P^{\mu} \right] +T_3(q^2) {\epsilon^*_{T } \cdot  q }\left[
       q^{\mu}-\frac{q^2}{m_B^2-m_{K_2^*}^2}P^{\mu}\right],\label{eq:BtoTformfactors-definition}
 \end{eqnarray}
where $q=P_B-P_2, P=P_B+P_2$. We also have the relation
$2m_{K_2^*}A_0(0)=(m_B+m_{K_2^*})A_1(0)-(m_B-m_{K_2^*})A_2(0)$ in
order to smear the pole at $q^2=0$.

Using the newly-studied light-cone distribution
amplitudes~\cite{Cheng:2010hn}, we have computed $B\to K_2^*$ form
factors~\cite{Wang:2010ni} in the perturbative QCD approach
(PQCD)~\cite{Keum:2000ph}. At the leading power, our predictions are found to obey the
nontrivial relations derived from the
large energy symmetry. This consistence may imply that the PQCD
results for the form factors are reliable and therefore suitable for
the study of the semileptonic $B$ decays. The recent computation in
light-cone QCD sum rules~\cite{Yang:2010qd} is also consistent with
ours. Results in the light-cone sum rules in conjunction with
$B$-meson wave functions~\cite{Wang:2010tz}, however, are too
large and thus not favored by the $B\to K_2^*\gamma$ data.  In our
work the $B\to K_2^*$ form factors are
$q^2$-distributed as~\cite{Wang:2010ni}
\begin{eqnarray}
 F(q^2)&=&\frac{F(0)}{(1-q^2/m_B^2)(1-a(q^2/m_B^2)+b(q^2/m_B^2)^2)},\label{eq:fit-B-T}
\end{eqnarray}
where $F$ denotes a generic form factor among $A_0,A_1,V,T_{1-3}$.
Neglecting  higher power corrections, $A_2$ is related to
$A_0$ and $A_1$ by
\begin{eqnarray}
  A_2(q^2)=\frac{m_B+m_{K_2^*}}{m_B^2-q^2}\left[(m_B+m_{K_2^*})A_1(q^2)-2m_{K_2^*}
  A_0(q^2)\right].
\end{eqnarray}
Numerical results for the $B\to K_2^*$ and $B_s\to f_2'(1525)$ form
factors at maximally recoil point and the two fitted parameters
$a,b$ are collected in table~\ref{Tab:formfactors}. The two kinds of
errors are from: decay constants of $B$ meson and shape parameter
$\omega_b$; $\Lambda_{\rm{QCD}}$, the scales $t$s and the threshold
resummation parameter $c$~\cite{Wang:2010ni}.

\begin{table}
\caption{$B\to K_2^*$ and $B_s\to f_2'(1525)$ form factors in the
PQCD approach. $F(0)$ denotes results at $q^2=0$ point while
$a,b$ are the parameters in the parametrization shown in
Eq.~\eqref{eq:fit-B-T}. The two kinds of errors are from: decay
constants of $B$ meson and shape parameter $\omega_b$;
$\Lambda_{\rm{QCD}}$, factorization scales $t$s and the threshold
resummation parameter $c$. }
 \label{Tab:formfactors}
 \begin{center}
 \begin{tabular}{ c c c ccc c c}
\hline \hline
 $F$       & $F(0)$  & $a$ &$b$                     \\
 \hline
\hline
 $V^{B K_{2}^*}$    & $0.21_{-0.04-0.03}^{+0.04+0.05}$
                    & $1.73_{-0.02-0.03}^{+0.02+0.05}$
                    & $0.66_{-0.05-0.01}^{+0.04+0.07}$  \\ \hline
 $A_0^{B K_{2}^*}$  & $0.18_{-0.03-0.03}^{+0.04+0.04}$
                    & $1.70_{-0.02-0.07}^{+0.00+0.05}$
                    & $0.64_{-0.06-0.10}^{+0.00+0.04}$ \\ \hline
 $A_1^{B K_{2}^*}$  & $0.13_{-0.02-0.02}^{+0.03+0.03}$
                    & $0.78_{-0.01-0.04}^{+0.01+0.05}$
                    & $-0.11_{-0.03-0.02}^{+0.02+0.04}$  \\ \hline
 $A_2^{B K_{2}^*}$  & $0.08_{-0.02-0.01}^{+0.02+0.02}$  &$--$  &$--$       \\
 \hline
 $T_1^{B K_{2}^*}$  & $0.17_{-0.03-0.03}^{+0.04+0.04}$
                    & $1.73_{-0.03-0.07}^{+0.00+0.05}$
                    & $0.69_{-0.08-0.11}^{+0.00+0.05}$  \\
 \hline
 $T_2^{B K_{2}^*}$  & $0.17_{-0.03-0.03}^{+0.03+0.04}$
                    & $0.79_{-0.04-0.09}^{+0.00+0.02}$
                    & $-0.06_{-0.10-0.16}^{+0.00+0.00}$    \\
 \hline
 $T_3^{B K_{2}^*}$  & $0.14_{-0.03-0.02}^{+0.03+0.03}$
                    & $1.61_{-0.00-0.04}^{+0.01+0.09}$
                    & $0.52_{-0.01-0.01}^{+0.05+0.15}$   \\
 \hline \hline
  $V^{B_s f_2'}$& $0.20_{-0.03-0.03}^{+0.04+0.05}$
                    & $1.75_{-0.00-0.03}^{+0.02+0.05}$
                    & $0.69_{-0.01-0.01}^{+0.05+0.08}$  \\
 \hline
  $A_0^{B_s f_2'}$ &$0.16_{-0.02-0.02}^{+0.03+0.03}$
                    & $1.69_{-0.01-0.03}^{+0.00+0.04}$
                    & $0.64_{-0.04-0.02}^{+0.00+0.01}$\\
 \hline
  $A_1^{B_s f_2'}$&$0.12_{-0.02-0.02}^{+0.02+0.03}$
                    & $0.80_{-0.00-0.03}^{+0.02+0.07}$
                    & $-0.11_{-0.00-0.00}^{+0.05+0.09}$ \\
 \hline
 $A_2^{B_s f_2'}$&$0.09_{-0.01-0.01}^{+0.02+0.02}$ &$--$  &$--$\\
 \hline
 $T_1^{B_s f_2'}$&$0.16_{-0.03-0.02}^{+0.03+0.04}$
                    & $1.75_{-0.00-0.05}^{+0.01+0.05}$
                    & $0.71_{-0.01-0.08}^{+0.03+0.06}$  \\
 \hline
 $T_2^{B_s f_2'}$&$0.16_{-0.03-0.02}^{+0.03+0.04}$
                    & $0.82_{-0.04-0.06}^{+0.00+0.04}$
                    & $-0.08_{-0.09-0.08}^{+0.00+0.03}$  \\
 \hline
  $T_3^{B_s f_2'}$&$0.13_{-0.02-0.02}^{+0.03+0.03}$
                    & $1.64_{-0.00-0.06}^{+0.02+0.06}$
                    & $0.57_{-0.01-0.09}^{+0.04+0.05}$   \\
 \hline\hline
 \end{tabular}
 \end{center}
 \end{table}
%


\section{Differential decay distributions and spin amplitudes}

In this section, we will discuss the kinematics of the quasi
four-body decay $B\to K_2^*(\to K\pi)l^+l^-$, define angular
observables  and collect the explicit formulas of helicity
amplitudes and/or transversity amplitudes.

\subsection{Differential decay distribution }

\begin{figure}\begin{center}
\includegraphics[scale=0.4]{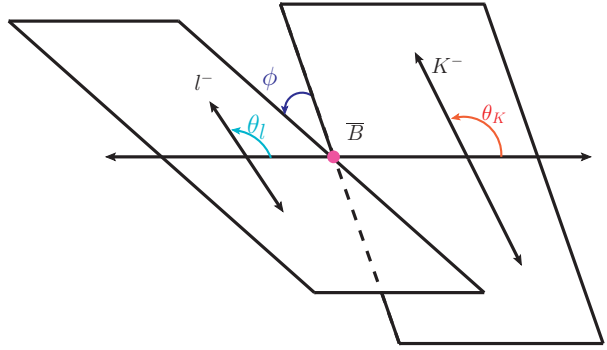}
\caption{Kinematics variables in the $\overline B\to  \bar K_2^*(\to
K^-\pi^+)l^+l^-$ process.  The moving direction of $K_2^*$ in  $B$
rest frame is chosen as the $z$ axis. The polar angle $\theta_K$
($\theta_l$) is defined as the angle between the flight direction of
$K^-$ ($\mu^-$) and the $z$ axis in the $K_2^*$ (lepton pair) rest
frame. The convention also applies to $B_s\to f_2'(\to
K^+K^-)l^+l^-$ transition.} \label{fig:angles}
\end{center}
\end{figure}

At the quark level, the decay amplitude for $b\to sl^+l^-$ is
expressed as
\begin{eqnarray}
 {\cal M}(b\to
 sl^+l^-)&=&\frac{G_F}{\sqrt2}\frac{\alpha_{\rm em}}{\pi}V_{tb}V_{ts}^*\times
 \left( \frac{C_9+C_{10}}{4}[\bar sb]_{V-A}[\bar ll]_{V+A}
 +\frac{C_9-C_{10}}{4}[\bar sb]_{V-A}[\bar ll]_{V-A}\right. \nonumber\\
 &&\left.+ C_{7L}m_b[\bar s i\sigma_{\mu\nu}
 (1+\gamma_5)b]\frac{q^\mu}{q^2}\times[\bar l \gamma^\nu l]+ C_{7R}m_b[\bar s i\sigma_{\mu\nu}
 (1-\gamma_5)b]\frac{q^\mu}{q^2}\times[\bar l \gamma^\nu
 l]\right),\label{eq:decay-amplitude-bsll-LR}
\end{eqnarray}
where $C_{7L}=C_7$ and $C_{7R}=\frac{m_s}{m_b}C_{7L}$ in the SM. Sandwiching
Eq.~\eqref{eq:decay-amplitude-bsll-LR} between the initial and final
states and replacing the spinor product $[\bar sb]$ by hadronic
matrix elements, one obtains the decay amplitude for hadronic
$B$ process. For the process under scrutiny in this work, the decay
observed in experiment is actually $B\to K_2^*(\to K\pi)l^+l^-$
which is a quasi four-body decay. The convention on the kinematics is illustrated in Fig.~\ref{fig:angles}. The moving
direction of $K_2^*$ in $B$ meson rest frame is chosen as $z$
axis. The polar angle $\theta_K$ ($\theta_l$) is defined as the
angle between the flight direction of $K^-$ ($\mu^-$) and the $z$
axis in $K_2^*$ (lepton pair) rest frame. $\phi$ is the angle
defined by decay planes of $K_2^*$ and the lepton pair.


Using the technique of helicity amplitudes described in the appendix B, we obtain the partial
decay width
\begin{eqnarray}
 \frac{d^4\Gamma}{dq^2d\cos\theta_K d\cos\theta_l d\phi}&=& \frac{3}{8}  |{\cal
 M}_B|^2,
\end{eqnarray}
with  the mass correction factor $\beta_l=\sqrt{1-4m_l^2/q^2}$. The
function $|{\cal M}_B|^2$ is decomposed into 11 terms
\begin{eqnarray}
 |{\cal M}_B|^2&=& \Big[I_1^c C^2 + 2I_1^s S^2+(I_2^c C^2 +2I_2^s S^2) \cos(2\theta_l) + 2I_3 S^2\sin^2\theta_l
 \cos(2\phi)+2\sqrt 2I_4CS\sin(2\theta_l)\cos\phi \nonumber\\
 && +2\sqrt 2I_5CS \sin(\theta_l) \cos\phi+2I_6S^2\cos\theta_l+2\sqrt 2I_7 CS\sin(\theta_l) \sin\phi\nonumber\\
 && +
 2\sqrt 2I_8CS\sin(2\theta_l)\sin\phi+2I_9 S^2\sin^2\theta_l
 \sin(2\phi)\Big],
\end{eqnarray}
with the angular coefficients
\begin{eqnarray}
 I_1^c&=&  (|A_{L0}|^2+|A_{R0}|^2)
 +8 \frac{m_l^2}{q^2}{\rm Re}[A_{L0}A^*_{R0} ]+4\frac{m_l^2}{q^2} |A_t|^2, \nonumber\\
 I_1^s&=&\frac{3}{4} [|A_{L\perp}|^2+|A_{L||}|^2+|A_{R\perp}|^2+|A_{R||}|^2  ]
 \left(1-\frac{4m_l^2}{3q^2}\right)+\frac{4m_l^2}{q^2} {\rm Re}[A_{L\perp}A_{R\perp}^*
 + A_{L||}A_{R||}^*],\nonumber\\
 I_2^c  &=& -\beta_l^2(  |A_{L0}|^2+ |A_{R0}|^2),\nonumber\\
 I_2^s  &=&
 \frac{1}{4}\beta_l^2(|A_{L\perp}|^2+|A_{L||}|^2+|A_{R\perp}|^2+|A_{R||}|^2),
 \nonumber\\
 I_3  &=&\frac{1}{2}\beta_l^2(|A_{L\perp}|^2-|A_{L||}|^2+|A_{R\perp}|^2-|A_{R||}|^2),\nonumber\\
 I_4
  &=& \frac{1}{\sqrt2}\beta_l^2
  [{\rm Re}(A_{L0}A_{L||}^*)+{\rm
  Re}(A_{R0}A_{R||}^*],\;\;\;\;\;\;\;\;
 I_5
  = \sqrt 2\beta_l
  [{\rm Re}(A_{L0}A_{L\perp}^*)-{\rm Re}(A_{R0}A_{R\perp}^*)],\nonumber\\
 I_6  &=& 2\beta_l
  [{\rm Re}(A_{L||}A^*_{L\perp})-{\rm
  Re}(A_{R||}A^*_{R\perp})],\;\;\;\;\;
 I_7
 = \sqrt2\beta_l
  [{\rm Im}(A_{L0}A^*_{L||})-{\rm Im}(A_{R0}A^*_{R||})],\nonumber\\
 I_8 &=& \frac{1}{\sqrt2}\beta_l^2
  [{\rm Im}(A_{L0}A^*_{L\perp})+{\rm
  Im}(A_{R0}A^*_{R\perp})],\;\;\;\;\;
 I_9
 =\beta_l^2
  [{\rm Im}(A_{L||}A^*_{L\perp})+{\rm
  Im}(A_{R||}A^*_{R\perp})].\label{eq:angularCoefficients}
\end{eqnarray}
$C=C(K_2^*)$ and $S=S(K_2^*)$ for $B\to K_2^*l^+l^-$. Without higher
order QCD corrections, $I_7$ is zero and $I_8,I_9$ are tiny in the
SM and the reason is that  only $C_9$ has an imaginary part.
In this sense these coefficients  can be chosen as an ideal window
to probe new physics signals.

The amplitudes $A_i$ are generated from the hadronic $B\to K_2^*V$
amplitudes ${\cal H}_i$ through $A_i=\sqrt{\frac{\sqrt \lambda  q^2 \beta_l}{3\cdot
32m_B^3\pi^3}{\cal B}(K_2^*\to K\pi)}{\cal H}_{i}$
\begin{eqnarray}
 A_{L0}
  &=& N_{K^*_2}  \frac{\sqrt{\lambda}}{\sqrt6 m_Bm_{K_2^*}}\frac{1}{2m_{K^*_2}\sqrt {q^2}}\left[ (C_9-C_{10})
[(m_B^2-m_{K^*_2}^2-q^2)(m_B+m_{K^*_2})A_1
 -\frac{\lambda}{m_B+m_{K^*_2}}A_2]\right.\nonumber\\
 &&\left. +  2m_b(C_{7L}-C_{7R})  [ (m_B^2+3m_{K_2^*}^2-q^2)T_2 -\frac{\lambda  }
 {m_B^2-m_{K_2^*}^2}T_3]\right],\nonumber\\
 A_{L\pm}
 &=& N_{K^*_2} \frac{\sqrt{\lambda}}{\sqrt8m_Bm_{K_2^*}}
  \left[ (C_9-C_{10}) [(m_B+m_{K^*_2})A_1\mp \frac{\sqrt \lambda}{m_B+m_{K^*_2}}V]\right.\nonumber\\
 &&\left.
 -\frac{2m_b(C_{7L}+C_{7R})}{q^2} (\pm\sqrt \lambda T_1)+\frac{2m_b(C_{7L}-C_{7R})}{q^2}
 (m_B^2-m_{K_2^*}^2)T_2\right],
 \nonumber\\
 A_{Lt}&=& N_{K^*_2}   \frac{\sqrt{\lambda}}{\sqrt 6m_Bm_{K_2^*}}
 (C_{9}-C_{10})\frac{\sqrt \lambda}{\sqrt {q^2}}A_0,
\end{eqnarray}
with  $N_{K_2^*}=[\frac{G_F^2 \alpha_{\rm em}^2}{3\cdot 2^{10}\pi^5
m_B^3}|V_{tb}V_{ts}^*|^2 q^2\lambda^{1/2}
\left(1-\frac{4m_l^2}{q^2}\right)^{1/2}{\cal B}(K_2^*\to
K\pi)]^{1/2}$. For convenience, we have introduced transversity
amplitudes as
\begin{eqnarray}
 A_{L\perp/||}&=&\frac{1}{\sqrt 2}(A_{L+}\mp A_{L-}),\nonumber\\
 A_{L\perp}&=& -\sqrt{2} \frac{\sqrt{\lambda}}{\sqrt8m_Bm_{K_2^*}}N_{K_2^*}\left[(C_9-C_{10})
 \frac{\sqrt \lambda V}{m_B+m_{K^*_2}}+\frac{2m_b(C_{7L}+C_{7R})}{q^2}\sqrt \lambda T_1\right],\nonumber\\
 A_{L||}&=& \sqrt{2}\frac{\sqrt{\lambda}}{\sqrt 8m_Bm_{K_2^*}} N_{K_2^*} \left[(C_9-C_{10}) (m_B+m_{K^*_2})A_1+\frac{2m_b(C_{7L}-C_{7R})}{q^2}(m_B^2-m_{K^*_2}^2)
 T_2 \right],
\end{eqnarray}
and the right-handed decay amplitudes are similar
\begin{eqnarray}
 A_{Ri}
  &=& A_{Li}|_{C_{10}\to -C_{10}}.
\end{eqnarray}
The combination of the timelike decay amplitude is used in the
differential distribution
\begin{eqnarray}
 A_t=A_{Rt}-A_{Lt}= 2N_{K^*_2} \frac{\sqrt{\lambda}}{\sqrt6m_Bm_{K_2^*}}C_{10}\frac{\sqrt \lambda}{\sqrt {q^2}}A_0.
\end{eqnarray}

\subsection{Dilepton spectrum distribution}

Integrating out the angles $\theta_l,\theta_K$ and $\phi$, we obtain
the dilepton mass spectrum
\begin{eqnarray}
 \frac{ d\Gamma}{dq^2}
 &=& \frac{1}{4} \left(3I_1^c+6I_1^s-I_2^c-2I_2^s\right),
\end{eqnarray}
and its expression in the massless limit
 \begin{eqnarray}
 \frac{ d\Gamma_i}{dq^2}
 &=&     (|A_{Li}|^2+|A_{Ri}|^2),
 \end{eqnarray}
with $i=0,\pm1$ or $i=0,\perp,||$.
After some manipulations in the appendix, the correspondence of the above equations and Eq.~(\ref{eq:AFBs-helicity}) with results in Ref.~\cite{Hatanaka:2009gb} can be shown.

\subsection{Polarization distribution}

The longitudinal polarization distribution for $\overline B\to \overline K_2^*l^+l^-$ is
defined as
\begin{eqnarray}
 \frac{df_L}{dq^2}\equiv
 \frac{d\Gamma_0}{dq^2}\Big/\frac{d\Gamma}{dq^2}=
 \frac{3I_1^c-I_2^c}{3I_1^c+6I_1^s-I_2^c-2I_2^s},
\end{eqnarray}
in which $\frac{d\Gamma_0}{dq^2}$ can be reduced into $I_1^c$ in the
case of $m_l=0$ since $I_1^c=-I_2^c$. The integrated polarization
fraction is given as
\begin{eqnarray}
  {f_L} \equiv \frac{\Gamma_0}{\Gamma}=\frac{\int
  dq^2\frac{d\Gamma_0}{dq^2}}{\int
  dq^2\frac{d\Gamma}{dq^2}}.
\end{eqnarray}

\subsection{Forward-backward asymmetry}

The differential forward-backward asymmetry of $\overline B\to \overline
K_2^*l^+l^-$ is defined by
\begin{eqnarray}
 \frac{d A_{FB}}{dq^2}&=&\left[\int_0^1 -\int_{-1}^0\right] d\cos\theta_l\frac{d^2\Gamma}{dq^2 d\cos\theta_l}
 =\frac{3}{4} I_6,\label{eq:AFBs-helicity}
\end{eqnarray}
while  the normalized differential FBA is
given by
\begin{eqnarray}
 \frac{\overline {dA_{FB}}}{dq^2}&=&\frac{\frac{d  A_{FB}}{dq^2}}
 { \frac{d\Gamma}{dq^2}}=
 \frac{3I_6}{3I_1^c+6I_1^s-I_2^c-2I_2^s}.
\end{eqnarray}

In the massless limit, we have
\begin{eqnarray}
 \frac{dA_{FB}}{dq^2}
 &=& \frac{ {\lambda}}{ 8m_B^2m_{K_2^*}^2}\frac{{\lambda}q^2 G_F^2\alpha_{\rm em}^2}{512\pi^5 m_B^3}
 \left|V_{tb}V_{ts}^*\right|^2
 {\rm Re}\Bigg[C_9C_{10}
 A_1 V+
 {C_{10}(C_{7L}+C_{7R})}\frac{m_b(m_B+m_{K_2^*})}{q^2}
 A_1 T_1
 \nonumber\\
 &&
 +{C_{10}(C_{7L}-C_{7R})} \frac{m_b(m_B-m_{K_2^*})}{q^2}  T_2
 V\Bigg].\label{eq:AFB-simplication-massless}
\end{eqnarray}
In the SM where $C_{7R}$ is small, the zero-crossing point $s_0$ of FBAs
is determined by the equation
\begin{eqnarray}
 C_9 A_1 (s_0)
 V (s_0)+C_{7L}\frac{m_b(m_B+m_{K_2^*})}{ s_0}A_1 (s_0)T_1 (s_0)+C_{7L}\frac{m_b(m_B-m_{K_2^*})}{s_0}T_2 (s_0)V (s_0)=0.
\end{eqnarray}


\subsection{Spin amplitudes and transverse asymmetries}

Using the above helicity/spin amplitudes, it is also possible to
construct several useful quantities which are ratios of different
amplitudes. The following ones, widely studied in the $B\to
K^*$ case, are stable against the uncertainties
from hadronic form factors
\begin{eqnarray}
 A_{T}^{(1)}&=& \frac{\Gamma_--\Gamma_+}{\Gamma_-+\Gamma_+}=
 \frac{-2 {\rm
 Re}(A_{||}A_{\perp}^*)}{|A_{\perp}|^2+|A_{||}|^2},\nonumber\\
 A_{T}^{(2)}&=&
 \frac{|A_{\perp}|^2-|A_{||}|^2}{|A_{\perp}|^2+|A_{||}|^2},\nonumber\\
 A_{T}^{(3)}&=& \frac{|A_{L0}A_{L||}^*+A_{R0}A_{R||}^*|}{\sqrt{|A_0|^2|A_{\perp}|^2}},\nonumber\\
 A_{T}^{(4)}&=&
 \frac{|A_{L0}A_{L\perp}^*-A_{R0}A_{R\perp}^*|}{|A_{L0}A_{L||}^*+A_{R0}A_{R||}^*|},
\end{eqnarray}
with the notation
\begin{eqnarray}
 A_iA_j^*= A_{Li}A_{Lj}^*+A_{Ri}A_{Rj}^*.
\end{eqnarray}
Due to the hierarchy in the SM $\Gamma_-\gg\Gamma_+$, $A^{(1)}_T$ is
close to 1 and therefore its deviation from 1 is more useful to reflect the
size of the NP effects.

\section{Two NP models}\label{sec:two-NP-models}

The $b\to sl^+l^-$ has a small branching fraction since the SM is
lack of tree level FCNC. It is not necessarily the same in
extensions. In this section we will briefly give an
overview of two NP models, which allow tree-level FCNC. Both of
these two models, vector-like quark model and family non-universal
$Z'$ model, do not introduce new type operators but instead modify
the Wilson coefficients $C_{9},C_{10}$. To achieve this goal, they
introduce an $SU(2)$ singlet down-type quark or a new gauge boson
$Z'$.

\subsection{Vector-like quark model: Z-mediated FCNCs}


In the vector-like quark model, the new $SU(2)_{L}$ singlet down
quarks $D_L$ and $D_R$ modify the Yukawa interaction sector
 \begin{eqnarray}
 {\cal L}_{Y} &=& \bar Q_L Y_D H d_R + h_D \bar Q_L  H D_R + m_D \bar D_L D_R + h.c.\,,
 \end{eqnarray}
where the flavor indices have been suppressed. $Q_L$ ($H$) is the
SU(2) quark (Higgs) doublet, $Y_D$ and $h_D$ are the Yuakwa
couplings and $m_D$ is the mass of exotic quark before electroweak
symmetry breaking. When the Higgs field acquires the vacuum
expectation value (VEV), the mass matrix of down type quark becomes
 \begin{eqnarray}
 m_d &=& \left(
           \begin{array}{ccc}
             Y^{ij}_D & | &{ h_D^i} \\
             -& - &- \\
             0 & |& m_D\\
           \end{array}
         \right)\,,
 \end{eqnarray}
which can be diagonalized by two unitary matrices
\begin{eqnarray}
 m^{\rm dia}_{d} &=& V^{L}_{D} m_d V^{R\dagger}_{D}\,. \label{eq:mass}
\end{eqnarray}

The SM coupling of $Z$-boson to fermions is flavor blind, and the
flavor in the process with exchange of $Z$-boson is conserved at
tree level. Unlikely  although the right-handed sector in the VQM is
the same as the SM, the new left-handed quark is $SU(2)_L$ singlet,
which carries the same hypercharge as right-handed particles.
Therefore the gauge interactions of left-handed down-type quarks
with $Z$-boson are given by
\begin{eqnarray}
 {\cal L}_Z&=& \bar Q_L \frac{g}{\cos\theta_W} (I_3-\sin^2\theta_W Q) Z\!\!\!\slash Q_L
 + \bar D_L \frac{g}{\cos\theta_W} (-\sin^2\theta_W Q) Z\!\!\!\slash
 D_L, 
\end{eqnarray}
where $g$ is the coupling constant of $SU(2)_{L}$, $\theta_W$ is the
Weinberg's angle, $P_{R(L)}=(1\pm\gamma_5)/2$. $I_3$ and $Q$ are
operators for the third component of  the weak isospin and the
electric charge, respectively.

%

Since the ratio $\xi_D$ of the coupling constants  deviates from unity: $
 \xi_D=-\sin^2\theta_W Q_D/(I_3^F-\sin^2\theta_W Q_F)$,
tree level FCNC can be induced after the diagonalization of the down-type quarks. 
For instance, the interaction for $b$-$s$-$Z$ in the VQM is given by
 \begin{eqnarray}
 {\cal L}_{b\to s}= \frac{g c^{s}_{L}\lambda_{sb} }{ \cos\theta_W} \bar s \gamma^{\mu}
  P_L b Z_{\mu} + h.c.,
 \label{eq:bsint}
 \end{eqnarray}
where $\lambda_{sb}$ is introduced as the new free parameter:
 \begin{eqnarray}
 \lambda_{sb}=(\xi_D -1)(V^{L}_D)_{sD}
(V^{L}_D)^*_{bD}\equiv|\lambda_{sb}|\exp\left(i\theta_{s}\right)\,.
\nonumber
 \end{eqnarray}


Using Eq.~(\ref{eq:bsint}), the effective Hamiltonian for $b\to
sl^+l^-$ mediated by $Z$-boson is found by
\begin{eqnarray}
{\cal H}^Z_{b\to s l^+l^-} &=& \frac{2G_F}{\sqrt{2}} \lambda_{sb}
c^s_L (\bar s b)_{V-A} \left[c^\ell_L (\bar\ell \ell)_{V-A}
+c^\ell_R (\bar \ell \ell)_{V+A} \right].
\end{eqnarray}
The Wilson coefficients $C_{9,10}$ are modified accordingly
\begin{eqnarray}
C_9^{\rm VLQ}  &=& C^{\rm SM}_9 - \frac{4\pi}{\alpha_{\rm
em}}\frac{\lambda_{sb} c^s_L (c^\ell_L+c_R^\ell)}{V^{*}_{ts}
V_{tb}}\,, \;\;\; C_{10}^{\rm VLQ} = C^{\rm SM}_{10} +
\frac{4\pi}{\alpha_{\rm em}}\frac{\lambda_{sb} c^s_L
 (c^\ell_L-c_R^\ell)}{V^{*}_{ts} V_{tb}}\,.\label{eq:NP-C9-C10-VLQ}
\end{eqnarray}
Making use of the experimental data of $b\to sl^+l^-$, our previous
work~\cite{Chen:2010aq} has placed a constraint on the new
coupling constant
\begin{eqnarray}
 |\lambda_{sb}|<1\times 10^{-3},
\end{eqnarray}
but its phase $\theta_s$ is less constrained.  In the following, we
shall see that the constraint can be improved by taking into account
the experimental data of the exclusive process $B\to K^*l^+l^-$.

\subsection{Family non-universal $Z'$ model}

The SM can be extended by including an additional $U(1)^{\prime}$
symmetry, and the currents can be given as following in a proper
gauge basis
\begin{eqnarray}
J_{Z^{\prime}}^{\mu}=g^{\prime}\sum_i \bar\psi_i
\gamma^{\mu}[\epsilon_i^{\psi_L}P_L+\epsilon_i^{\psi_R}P_R]\psi_i,
\label{eq:JZprime}
\end{eqnarray}
where $i$ is the family index and $\psi$ labels the fermions (up- or
down-type quarks, or charged or neutral leptons).  According to some
string construction or GUT models such as $E_6$, it is possible to
have family non-universal $Z^{\prime}$ couplings, namely, even
though $\epsilon_i^{L,R}$ are diagonal the gauge couplings are not
family universal. After rotating to the physical basis, FCNCs
generally appear at tree level in both LH and RH sectors.
Explicitly,
\begin{eqnarray}
B^{\psi_L}=V_{\psi_L}\epsilon^{\psi_L}V_{\psi_L}^{\dagger},\;\;\;\;\;
B^{\psi_R}=V_{\psi_R}\epsilon^{\psi_R}V_{\psi_R}^{\dagger}.
\end{eqnarray}
Moreover, these couplings may contain CP-violating phases beyond
that of the SM.

The Lagrangian of $Z^{\prime}\bar bs$ couplings is given as
\begin{eqnarray}
{\cal L}_{\rm{FCNC}}^{Z^{\prime}}=-g^{\prime}(B_{sb}^L\bar
s_L\gamma_{\mu}b_L + B_{sb}^R\bar s_R\gamma_{\mu}b_R)Z^{\prime\mu} +
{\rm h.c.} . \label{eq:HamiltonZ}
\end{eqnarray}
It contributes to the $b\to s\ell^+\ell^-$ decay at tree level
with the effective Hamiltonian
\begin{eqnarray}
 {\cal H}_{\rm{eff}}^{Z^{\prime}}=
\frac{8G_F}{\sqrt{2}} (\rho_{sb}^L\bar s_L\gamma_{\mu}b_L +
\rho_{sb}^R\bar s_R\gamma_{\mu}b_R) (\rho_{ll}^L\bar
\ell_L\gamma^{\mu}\ell_L +\rho_{ll}^R\bar \ell_R\gamma^{\mu}\ell_R)
~, \label{eq:HamiltonZprime}
\end{eqnarray}
where
\begin{eqnarray}
\rho_{ff'}^{L,R} \equiv \frac{g'M_Z}{gM_{Z'}} B_{ff'}^{L,R}
\end{eqnarray}
with   the coupling  $g$ associated with the $SU(2)_L$ group in the
SM.
In this paper we shall not take the renormalization group running
effects due to these new contributions into consideration because
they are expected to be small. For the couplings are all unknown,
one can see from Eq. (\ref{eq:HamiltonZprime}) that there are many
free parameters here. For the purpose of illustration and to avoid
too many free parameters, we put the constraint that the FCNC
couplings of the $Z^{\prime}$ and quarks only occur in the
left-handed sector. Therefore, $\rho_{sb}^R=0$, and the effects of
the $Z^{\prime}$ FCNC currents simply modify the Wilson coefficients
$C_9$ and $C_{10}$ in Eq.~(\ref{eq:Hamiltonian}).  We denote these
two modified Wilson coefficients by $C_9^{Z^{\prime}}$ and
$C_{10}^{Z^{\prime}}$, respectively. More explicitly,
\begin{eqnarray}
  C_9^{Z^{\prime}} &=& C_9^{} -\frac{4\pi}{\alpha_{\rm em}}  {\frac{  \rho_{sb}^L
(\rho_{ll}^L+\rho_{ll}^R)}{
 V_{tb}V^*_{ts} }},\;\;\;
  C_{10}^{Z^{\prime}} =C_{10}+\frac{4\pi}{\alpha_{\rm em}}{  \frac{ \rho_{sb}^L(\rho_{ll}^L-\rho_{ll}^R) }{
  V_{tb}V^*_{ts} }}.
  \label{eq:wilsonZprime}
\end{eqnarray}

Compared with the Wilson coefficients in the vector-like quark model
in Eq.~\eqref{eq:NP-C9-C10-VLQ}, we can see that the $Z'$
contributions in Eq.~\eqref{eq:wilsonZprime} have similar forms and
the correspondence lies in the coupling constants
\begin{eqnarray}
 \lambda_{sb} c_L^s \to \rho_{sb}^L,\;\;\;c^l_{L,R}\to
 \rho_{ll}^{L,R}.
\end{eqnarray}
However the number of free parameters is increased from 2 to 4 since
$c^l_{L,R}$ in the VQM is the same as the SM.

\section{Phenomenological analysis}

In this section, we will present our theoretical results in the SM,
give an update of the constraints in the above two NP models and
investigate their effects on $B\to K_2^*\mu^+\mu^-$ and $B_s\to
f_2'\mu^+\mu^-$. For convenience, branching ratios of $K_2^*$
and $f_2'$ decays into $K\pi$ and $K\bar K$ will not be taken into
account in the numerical analysis.

\subsection{SM predictions}

\begin{figure}\begin{center}
\includegraphics[scale=0.7]{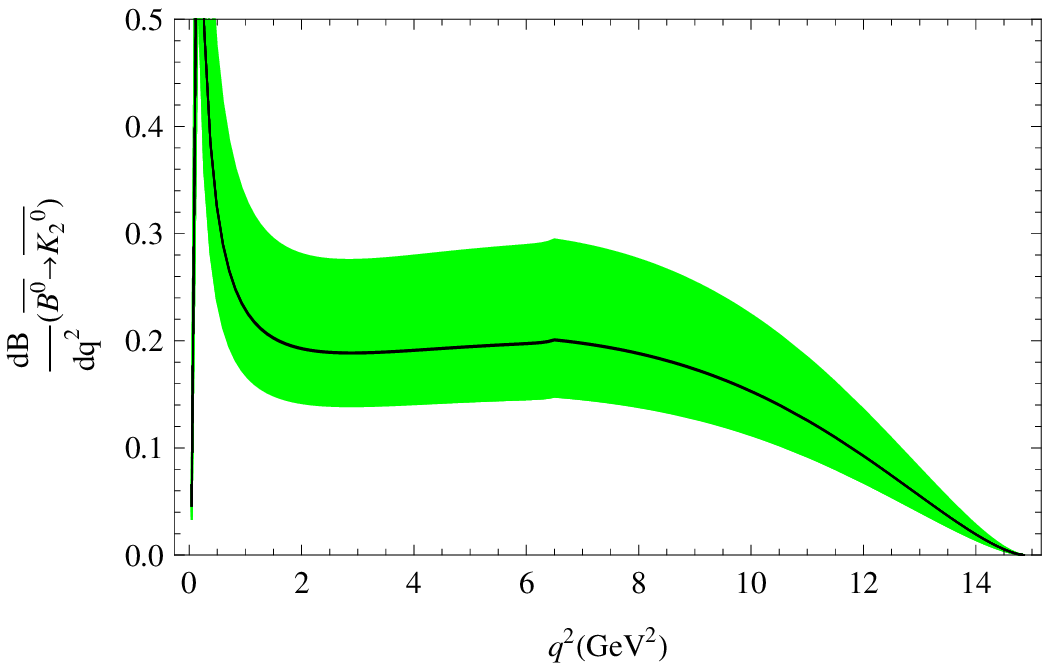} \hspace{0.8cm}
\includegraphics[scale=0.7]{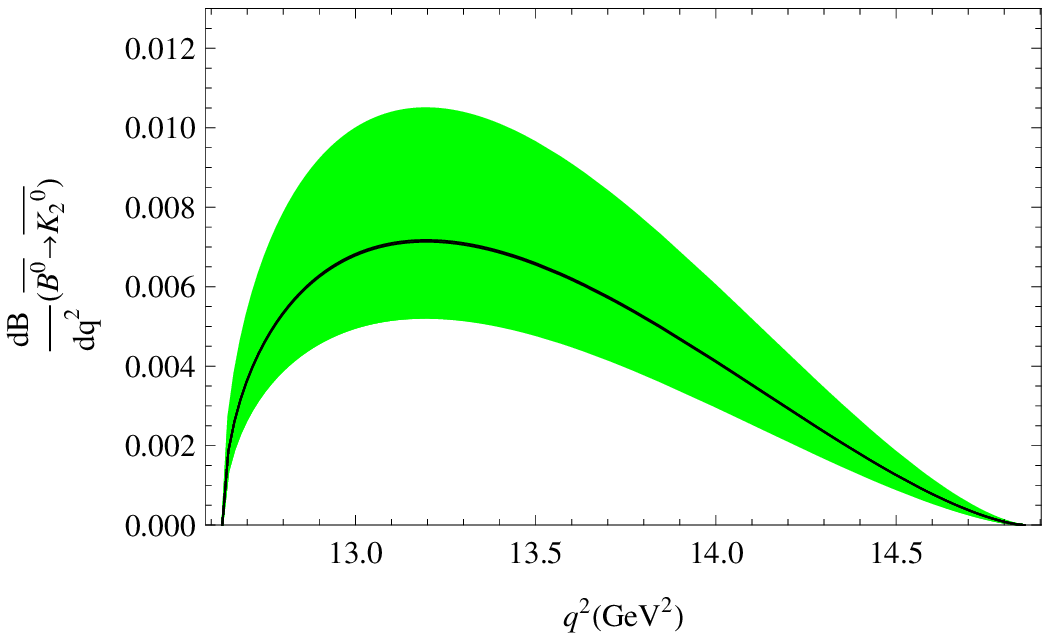} \\
\includegraphics[scale=0.7]{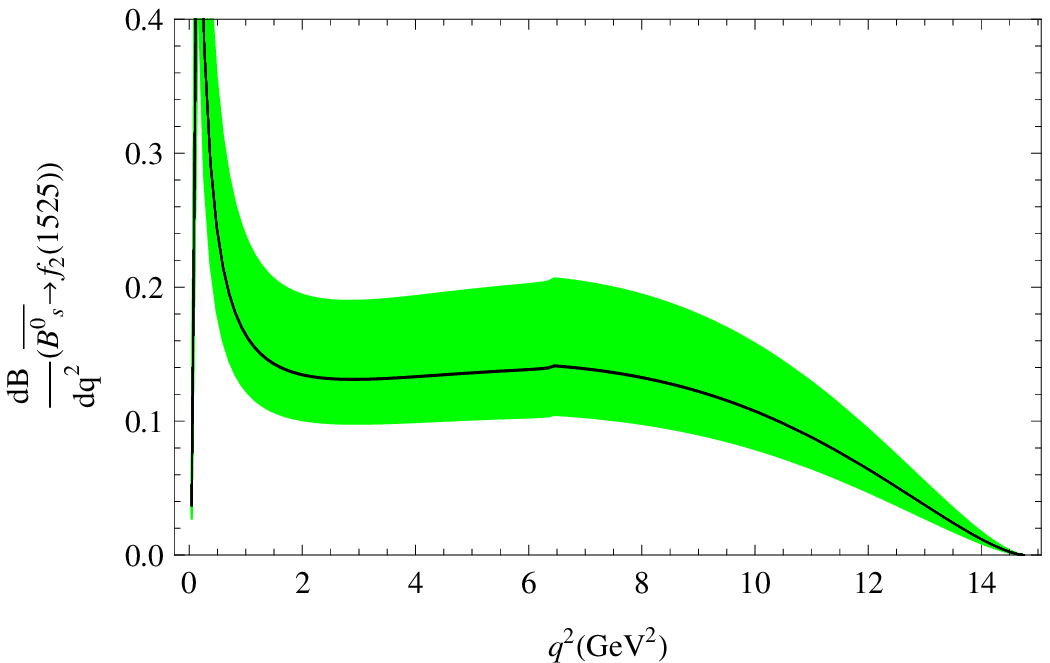} \hspace{0.8cm}
\includegraphics[scale=0.7]{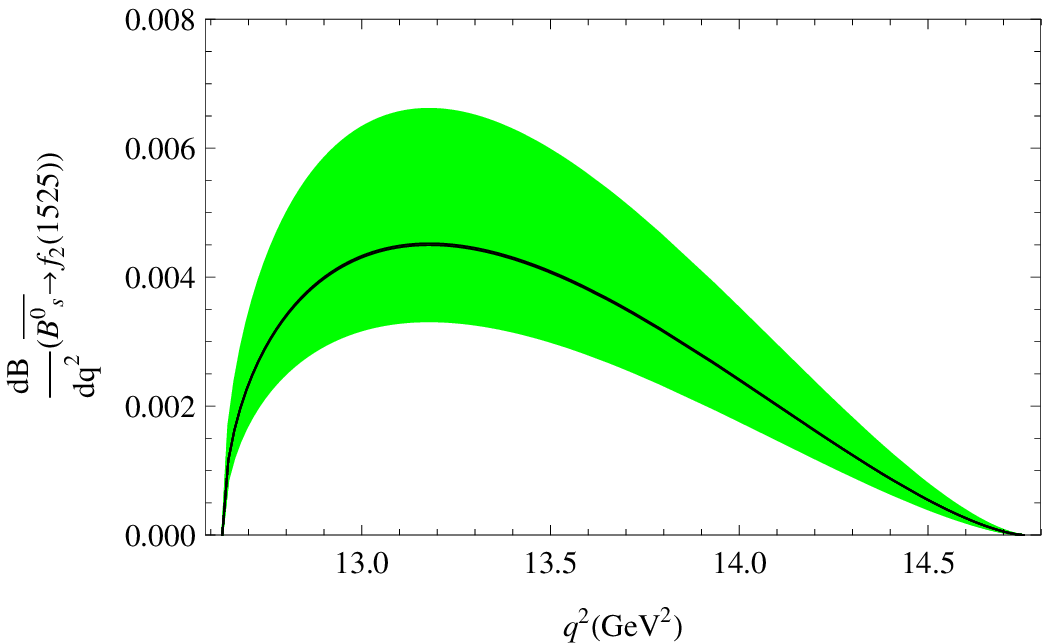} 
\caption{Differential branching ratios  of $B\to K_2^*l^+l^-$ (upper) and
$B_s\to f_2'l^+l^-$ (lower) ( in units of $10^{-7}$): the left panel  for $l=\mu$ and the right
panel for $l=\tau$.
}\label{dia:differential-BR-SM}
\end{center}
\end{figure}

With the $B\to K_2^*$ form factors computed in the PQCD approach~\cite{Wang:2010ni},
the BR, zero-crossing point of FBAs and polarization fractions are
predicted as
\begin{eqnarray}
 {\cal B}(B\to K_2^{*}\mu^+\mu^-)&=& (2.5^{+1.6}_{-1.1})\times 10^{-7},\nonumber\\
 f_L(B\to K_2^{*}\mu^+\mu^-)&=& (66.6\pm0.4)\%,\nonumber\\
 s_0(B\to K_2^{*}\mu^+\mu^-)&=& (3.49\pm0.04){\rm GeV}^2,\nonumber\\
 {\cal B}(B\to K_2^{*}\tau^+\tau^-)&=& (9.6^{+6.2}_{-4.5})\times 10^{-10},\nonumber\\
 f_L(B\to K_2^{*}\tau^+\tau^-)&=& (57.2\pm0.7)\%.
\end{eqnarray}
The errors are from the form factors, namely, from the $B$ meson
wave functions and the PQCD systematic parameters. Most of the
uncertainties from form factors will cancel in the polarization
fractions and the zero-crossing point $s_0$.  Similarly results for 
$B_s\to f_2l^+l^-$ are given as
\begin{eqnarray}
 {\cal B}(B_s\to f_2'\mu^+\mu^-)&=& (1.8^{+1.1}_{-0.7})\times 10^{-7},\nonumber\\
 f_L(B_s\to f_2' \mu^+\mu^-)&=& (63.2\pm0.7)\%,\nonumber\\
 s_0(B_s\to f_2' \mu^+\mu^-)&=& (3.53\pm0.03){\rm GeV}^2,\nonumber\\
 {\cal B}(B_s\to f_2' \tau^+\tau^-)&=& (5.8^{+3.7}_{-2.1})\times 10^{-10},\nonumber\\
 f_L(B_s\to f_2' \tau^+\tau^-)&=& (53.9\pm0.4)\%.
\end{eqnarray}
We also show the $q^2$-dependence of their differential branching
ratios (in units of $10^{-7}$) in Fig.~\ref{dia:differential-BR-SM}.

Charm-loop effects, due to the large Wilson coefficient and the
large CKM matrix element, might introduce important effects. In a
very recent work~\cite{Khodjamirian:2010vf}, the authors have
adopted QCD sum rules to investigate both factorizable diagrams
and nonfactorizable diagrams. Their results up to the region
$q^2=m_{J/\psi}^2$ are parameterized in the following form,
\begin{eqnarray}
 \Delta C_9^{(i)B\to K^*}(q^2)= \frac{r_1^{(i)}\left(1-\frac{\bar q^2}{q^2}\right)
 +\Delta C_9^{(i)}(\bar q^2)\frac{\bar q^2}{q^2}}{1+r_2^{(i)}\frac{\bar
 q^2-q^2}{m_{J/\psi}^2}},
\end{eqnarray}
where the three results correspond to
different Lorentz structures: $i=1,2,3$ for terms containing $V$,
$A_1$ and $A_2$ respectively. The numerical results are quoted as
follows
\begin{eqnarray}
 \Delta C_9^{(1)}(\bar q^2)=0.72^{+0.57}_{-0.37},\;\;\;
 r_1^{(1)}=0.10,\;\;
 r_2^{(1)}=1.13,\nonumber\\
 \Delta C_9^{(2)}(\bar q^2)=0.76^{+0.70}_{-0.41},\;\;\;
 r_1^{(2)}=0.09,\;\;
 r_2^{(2)}=1.12,\nonumber\\
 \Delta C_9^{(3)}(\bar q^2)=1.11^{+1.14}_{-0.70},\;\;\;
 r_1^{(3)}=0.06,\;\;
 r_2^{(3)}=1.05.
\end{eqnarray}
It should be pointed out that not all charm-loop effects in $B\to
K^*_2l^+l^-$ are the same as the ones in $B\to K^*l^+l^-$. Among
various diagrams the factorizable contributions, which can be simply
incorporated into $C_9$ given in Eq.~\eqref{eq:C7C9eff}, are the
same. The nonfactorizable ones are more subtle. In particular the
light-cone sum rules (LCSR) with $B$-meson distribution amplitudes
are adopted in Ref.~\cite{Khodjamirian:2010vf}, in which intermediate states like $K^*$ are picked up
as the ground state. The
generalization is not straightforward to the case of $K_2^*$ since in
this approach states below $K_2^*$ may contribute in a substantial manner. However in
another viewpoint, i.e.  the conventional LCSR, they
may be related. In our previous work we have shown that the
light-cone distribution amplitudes of $K_2^*$ is similar with $K^*$
in the dominant region of the PQCD approach. If it were also the
same in the conventional LCSR, one may expect that the charm-loop
effects in the processes under scrutiny  have
similar behaviors with the ones in $B\to K^*l^+l^-$. Therefore as
the first step to proceed, we will use their results to estimate the sensitivity in our following analysis
and to be conservative, we use
\begin{eqnarray}
 \Delta C_9^{(i)B\to K_2^*}(\bar q^2)=(1\pm1) \Delta C_9^{(i)B\to
 K^*}(q^2)
\end{eqnarray}
in the region of $1 {\rm GeV}^2<q^2<6 {\rm GeV}^2$.  The
central values for $q^2$-dependent parameters will be used for
simplicity and in this procedure, the factorizable corrections to
$C_9$ given in Eq.~\eqref{eq:C7C9eff} should be set to 0 to avoid
double counting.

\begin{figure}\begin{center}
\includegraphics[scale=0.7]{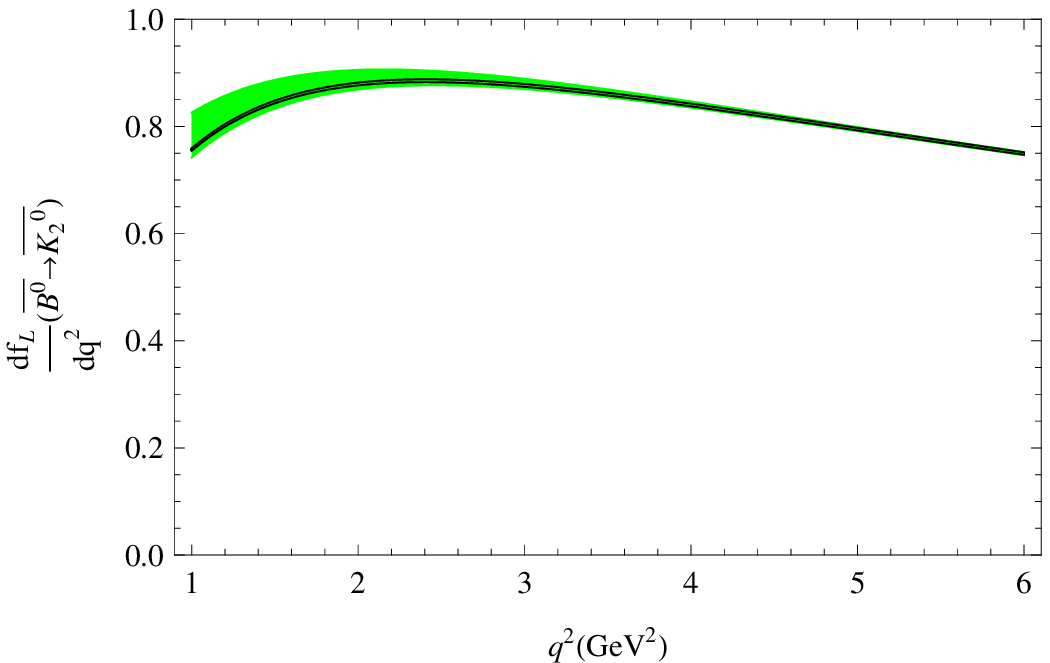} 
\hspace{1.cm}
\includegraphics[scale=0.7]{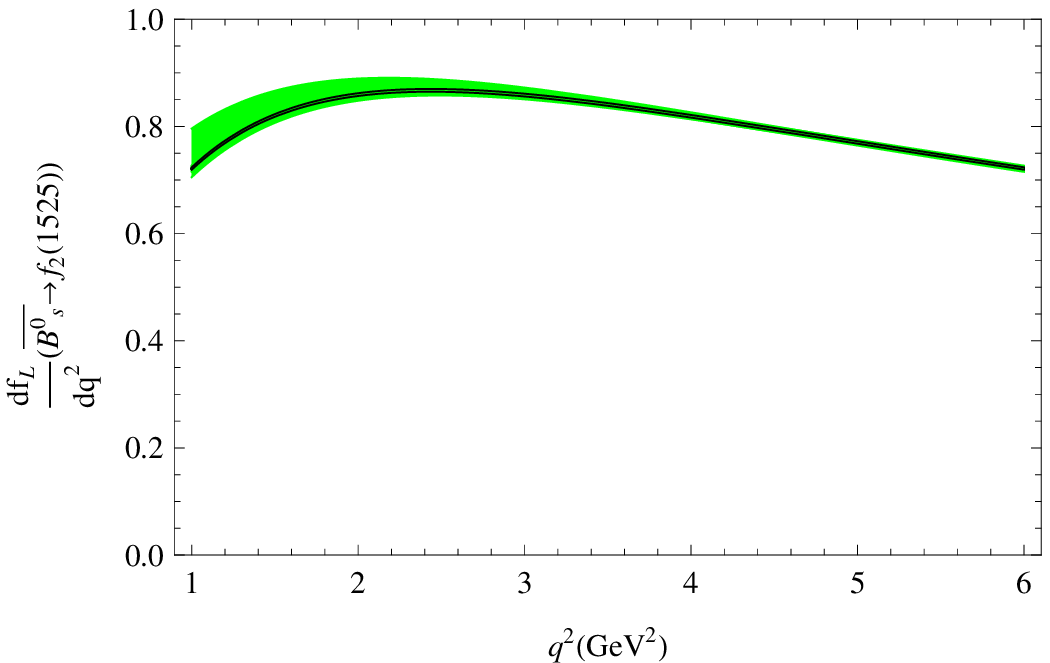} 
\\\caption{Differential polarization fractions $\frac{df_L}{dq^2}$ of
$B\to K_2^*l^+l^-$ (the left panel) and  $B_s\to f_2l^+l^-$ (the
right panel). }\label{dia:differential-fL-SM}
\end{center}
\end{figure}

\begin{figure}\begin{center}
\includegraphics[scale=0.7]{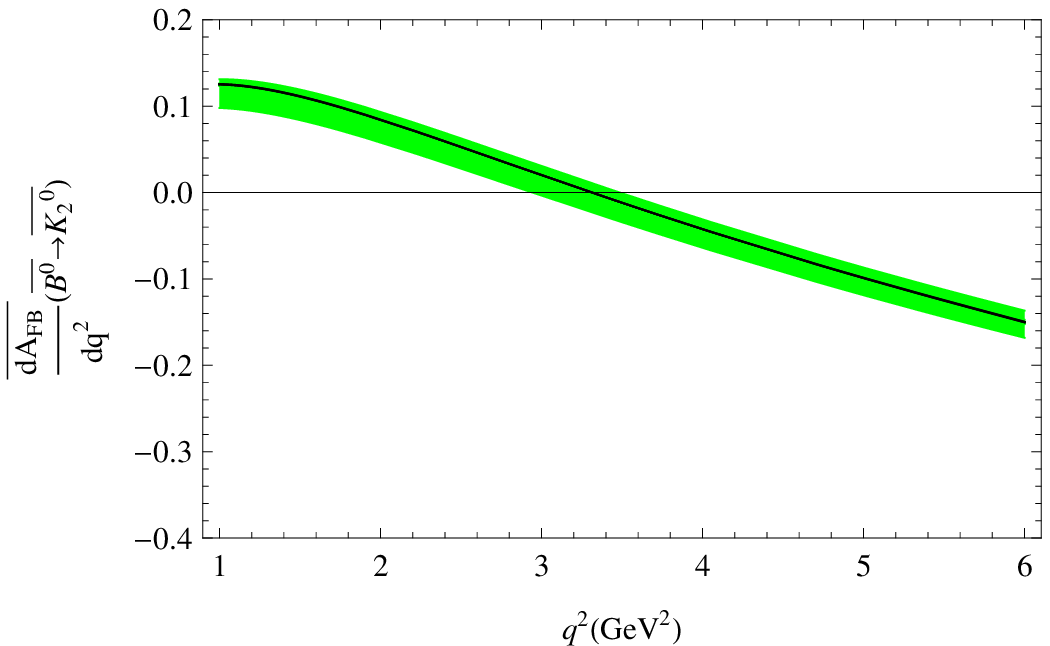} 
\hspace{1.cm}
\includegraphics[scale=0.7]{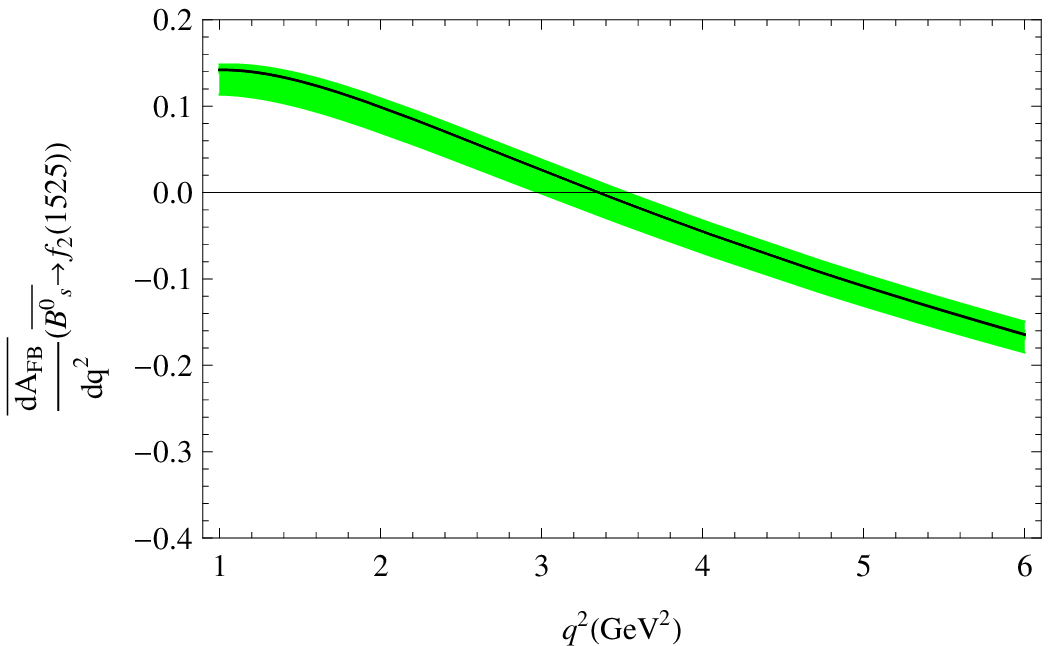} 
\caption{Similar with Fig.~\ref{dia:differential-fL-SM} but for
forward-backward asymmetries $
 \frac{\overline{d  A_{FB}}}{dq^2}$. }\label{dia:differential-AFB-SM}
\end{center}
\end{figure}

With the above strategy, our theoretical predictions are changed to
\begin{eqnarray}
 f_L(B\to K_2^{*}\mu^+\mu^-)&=&(66.6^{+1.4}_{-0.7})\%,\nonumber\\
 s_0(B\to K_2^{*}\mu^+\mu^-)&=&(3.49^{+0.19}_{-0.39}){\rm GeV}^2,\nonumber\\
 f_L(B_s\to f_2'\mu^+\mu^-)&=& (63.2^{+1.5}_{-0.9})\%,\nonumber\\
 s_0(B_s\to f_2'\mu^+\mu^-)&=& (3.53^{+0.19}_{-0.39}){\rm GeV}^2.
\end{eqnarray}
The uncertainties in  the zero-crossing point of FBAs are enlarged to
$0.4{\rm GeV}^2$. We also
show the $q^2$-dependence of the differential polarization in
Fig.~\ref{dia:differential-fL-SM} and the normalized
forward-backward asymmetries in Fig.~\ref{dia:differential-AFB-SM}.

As a parallel way, the BR of $B\to K_2^*l^+l^-$ can also be
estimated by making use of the data of radiative $B\to
K^*(K_2^*)\gamma$ decays~\cite{HFAG}
\begin{eqnarray}
 {\cal B}(\bar B^0\to K_2^*\gamma)&=& (12.4\pm2.4)\times
 10^{-6},\nonumber\\
 {\cal B}(\bar B^0\to K^*\gamma)&=& (43.3\pm1.5)\times
 10^{-6}.
\end{eqnarray}
The ratio of the above BRs $R\equiv  \frac{{\cal B}(K_2^*)}{{\cal
B}(K^*)}=0.29\pm0.06 $ and the measured data of $B\to K^*l^+l^-$
shown in Tab~\ref{tab:inputs-data} give the implication
\begin{eqnarray}
 {\cal B}_{\rm exp}(B^0\to K_2^{*0}l^+l^-)&=& (3.1\pm0.7)\times
 10^{-7},
\end{eqnarray}
which are remarkably consistent with our theoretical predictions
within uncertainties.

When the large energy symmetry is exploited,
the seven $B\to K_2^*$ form factors  can be reduced into two independent ones $\zeta_\perp$ and $\zeta_{||}$. Based on these nontrivial relations, 
Ref.~\cite{Hatanaka:2009gb} has used the experimental data of $B\to K_2^*\gamma$ to extract $\zeta_\perp$. With the assumption of a similar size for $\zeta_{||}$, the authors also  estimated the branching ratio and forward-backward asymmetries of $B\to K_2^*l^+l^-$.  Explicitly  they have employed 
\begin{eqnarray}
 \zeta_\perp=0.27\pm0.03^{+0.00}_{-0.01},\;\;\; 0.8\zeta_\perp< \zeta_{||}<1.2\zeta_\perp,
\end{eqnarray}
which are comparable with our results~\cite{Wang:2010ni}
\begin{eqnarray}
\zeta_\perp=(0.29\pm0.09),\;\;\;\zeta_{||}=(0.26\pm0.10).
\end{eqnarray}
As a consequence, the predicted results of BR, forward-backward asymmetries and polarizations are compatible with each other.

\begin{figure}\begin{center}
\includegraphics[scale=0.7]{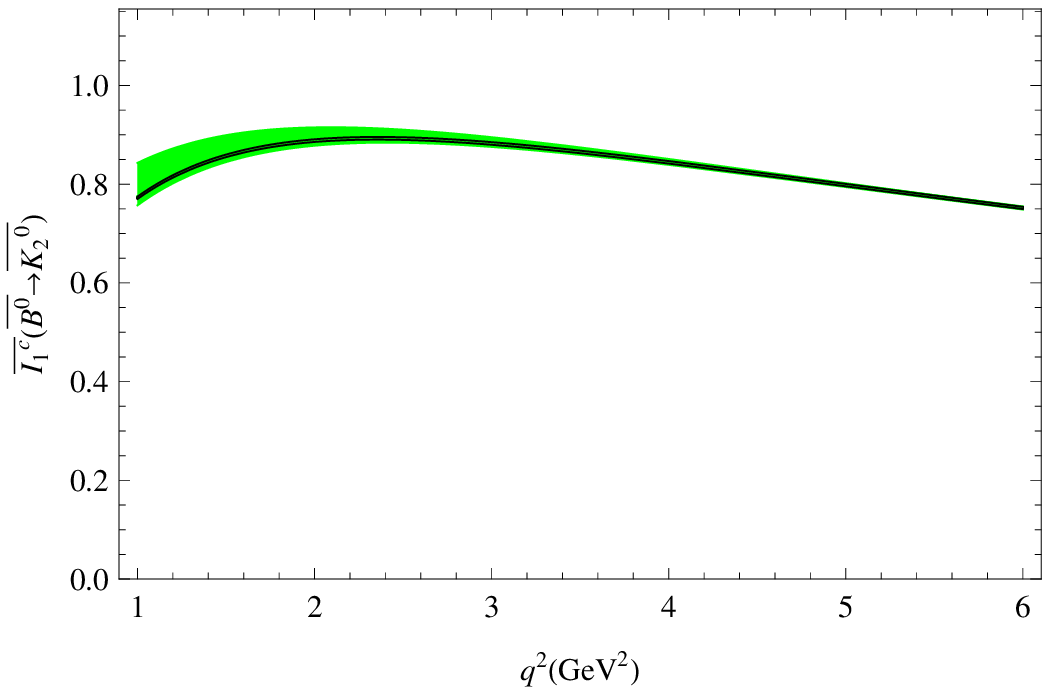} 
\hspace{1.cm}
\includegraphics[scale=0.7]{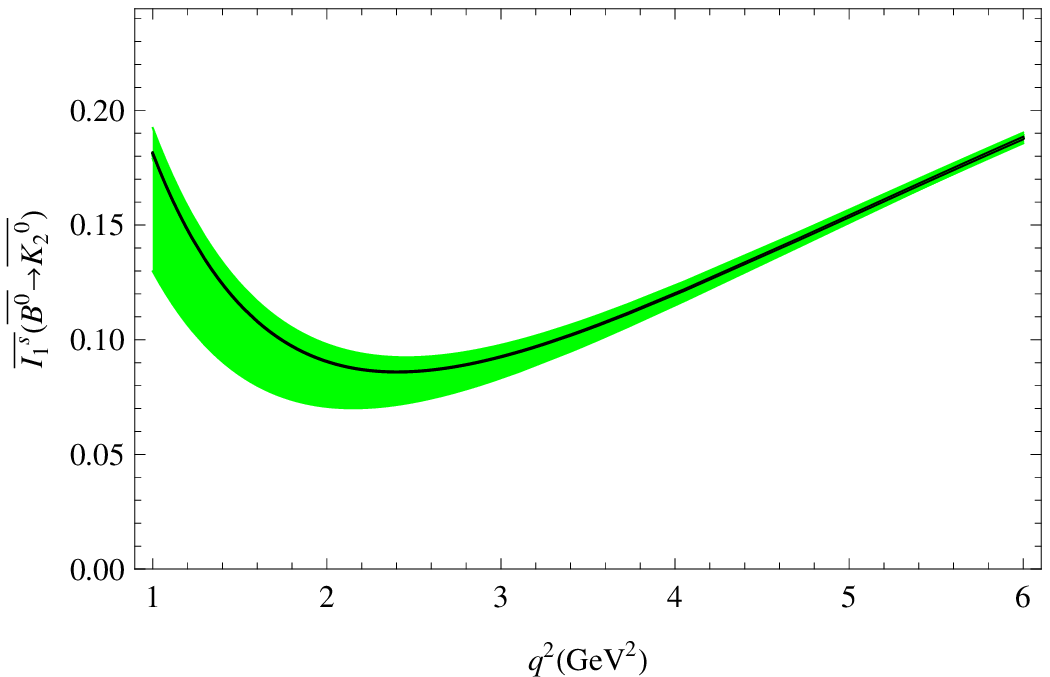} \\
\includegraphics[scale=0.7]{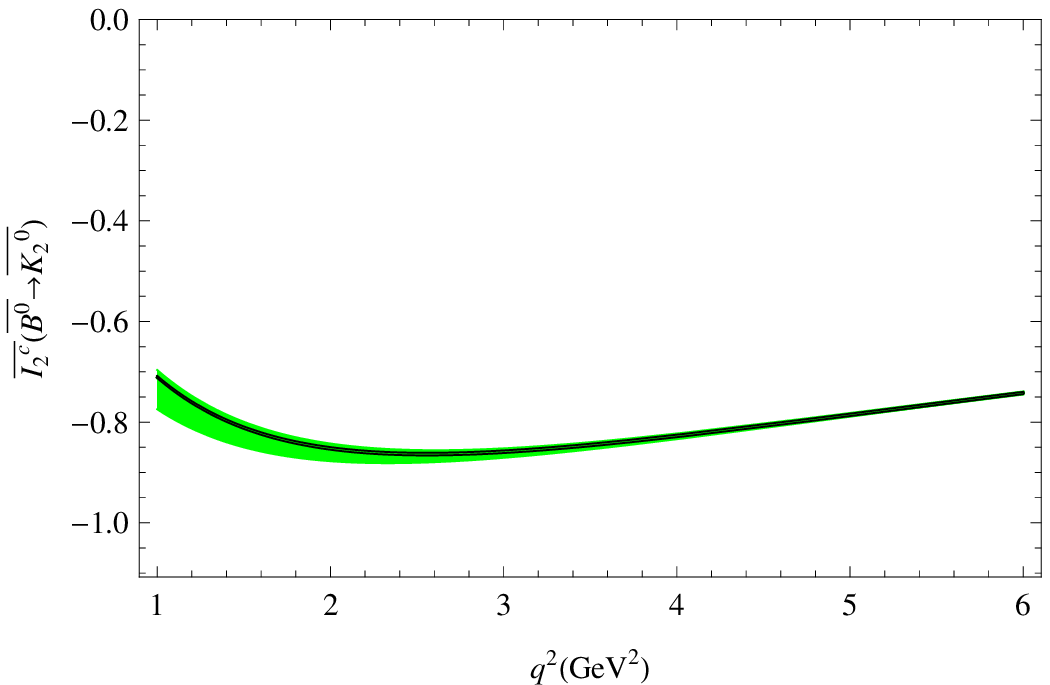} 
\hspace{1.cm}
\includegraphics[scale=0.7]{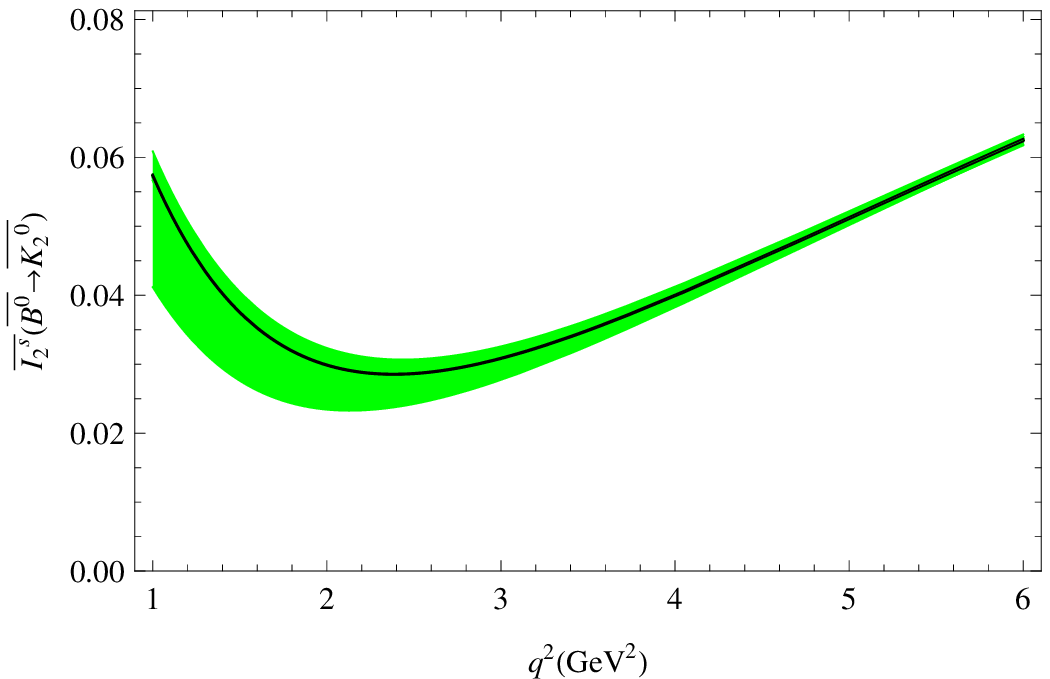} \\
\includegraphics[scale=0.7]{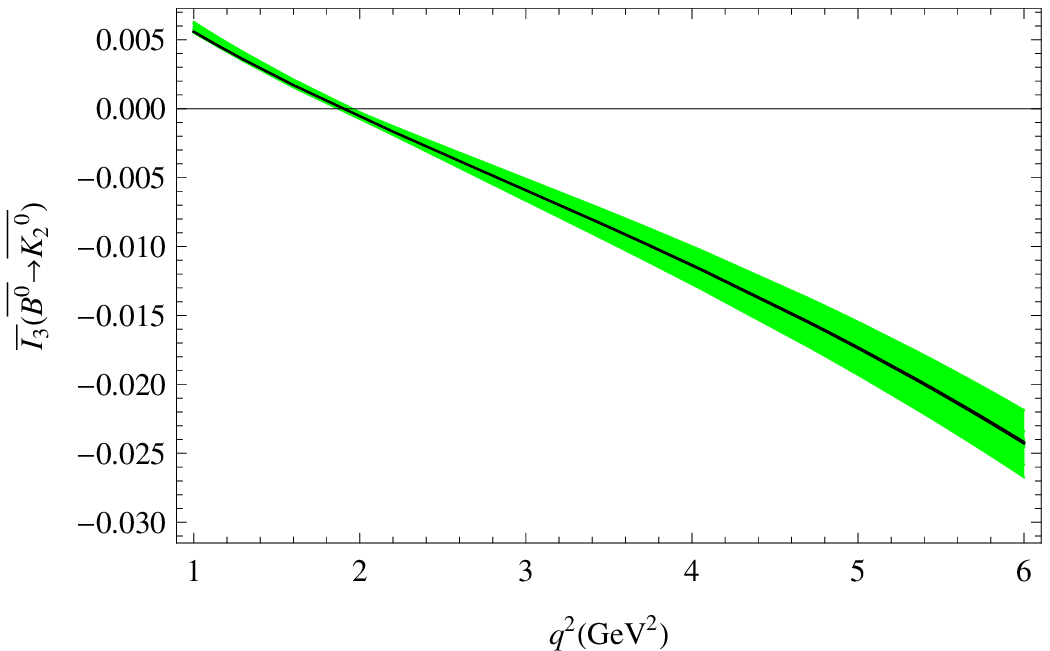} 
\hspace{1.cm}
\includegraphics[scale=0.7]{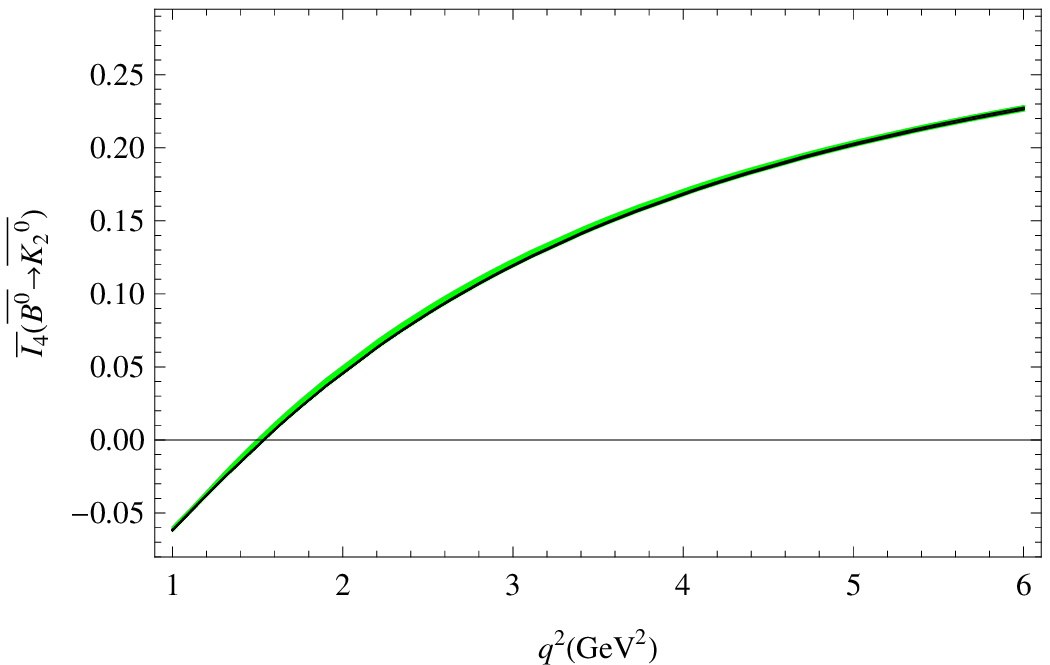} \\
\includegraphics[scale=0.7]{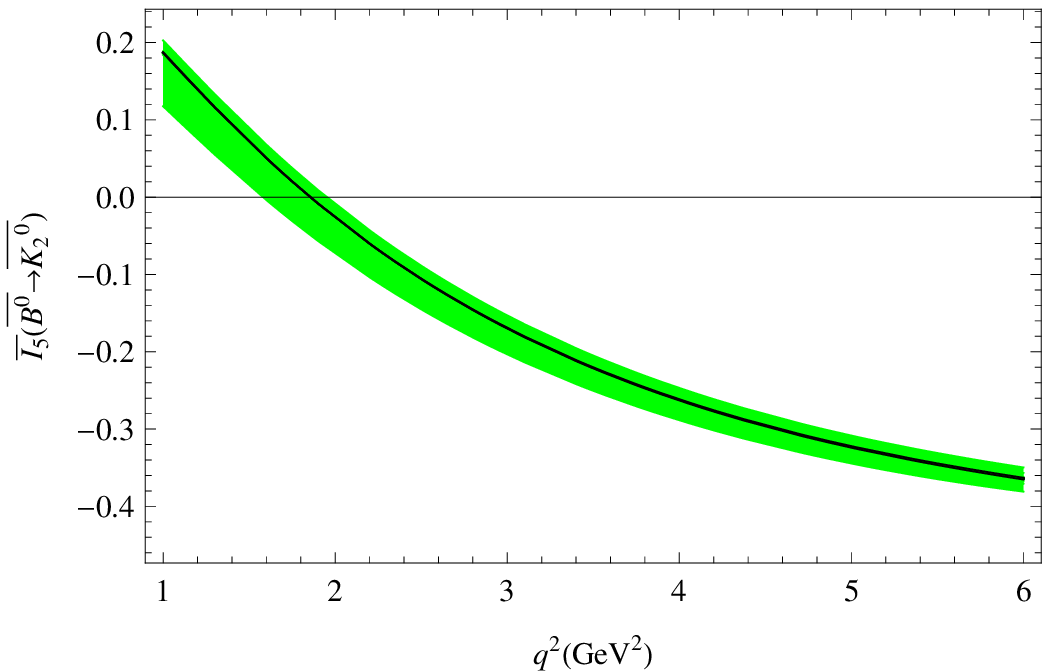} 
\hspace{1.cm}
\includegraphics[scale=0.7]{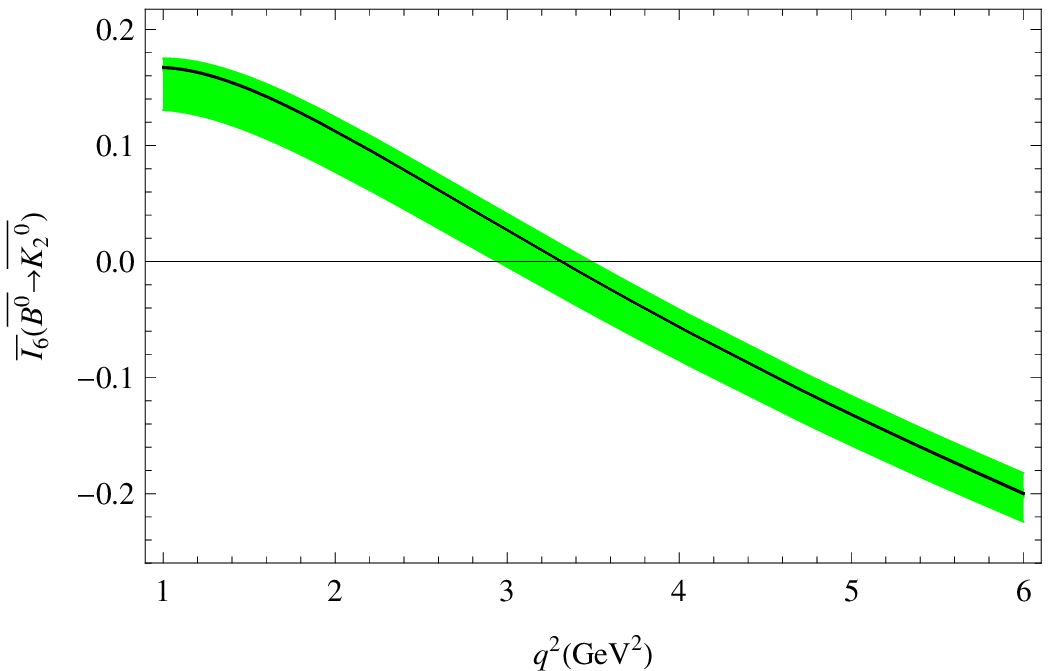} \\
\caption{Angular coefficients $\bar I_i$ for $B\to
K_2^*\mu^+\mu^-$}\label{Fig:angular-coefficients-B-K2-mumu}
\end{center}
\end{figure}

\begin{figure}\begin{center}
\includegraphics[scale=0.7]{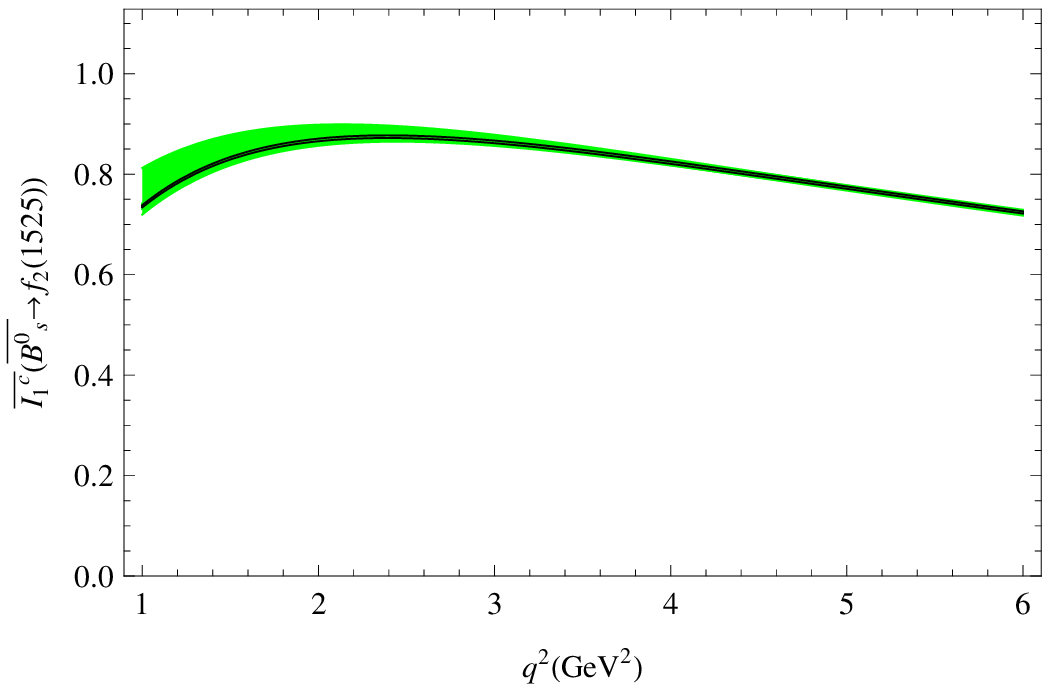} 
\hspace{1.cm}
\includegraphics[scale=0.7]{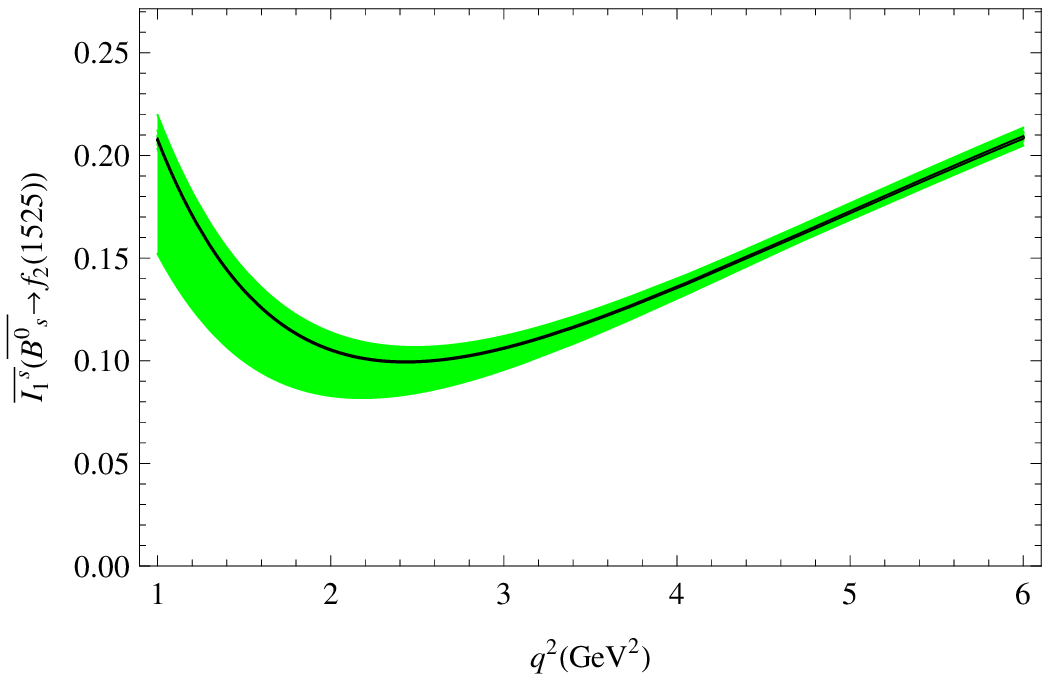} \\
\includegraphics[scale=0.7]{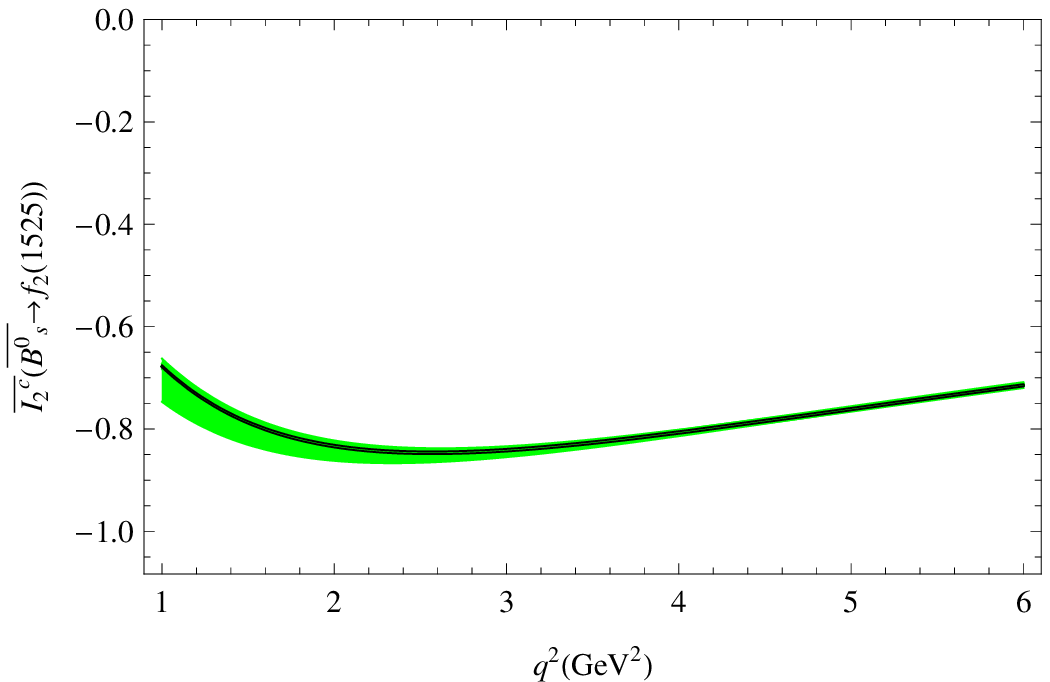} 
\hspace{1.cm}
\includegraphics[scale=0.7]{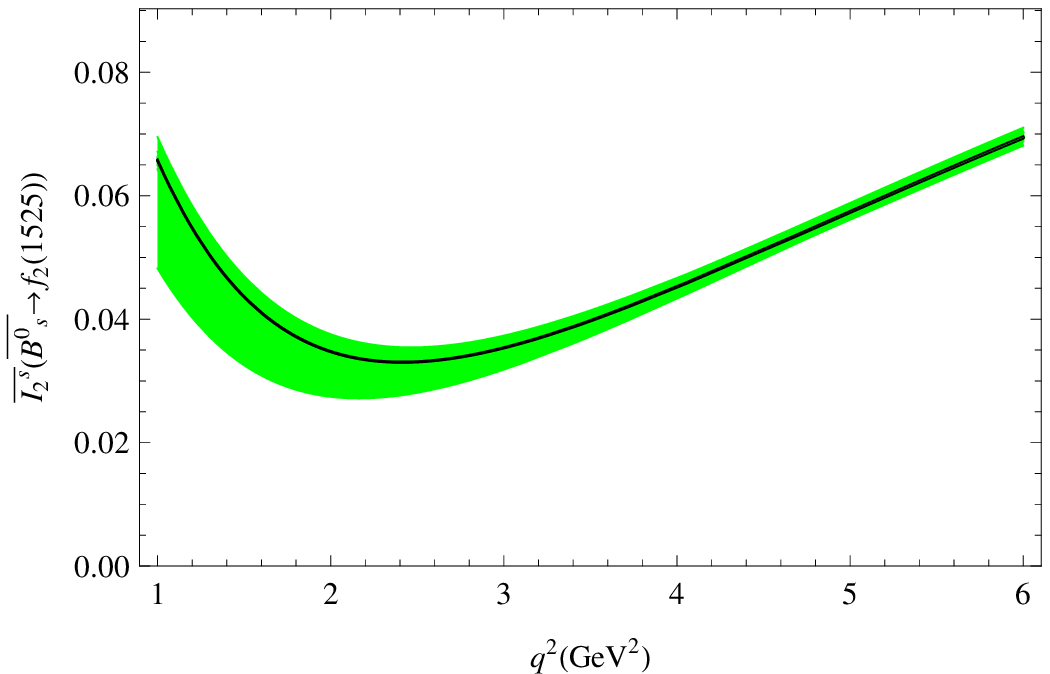} \\
\includegraphics[scale=0.7]{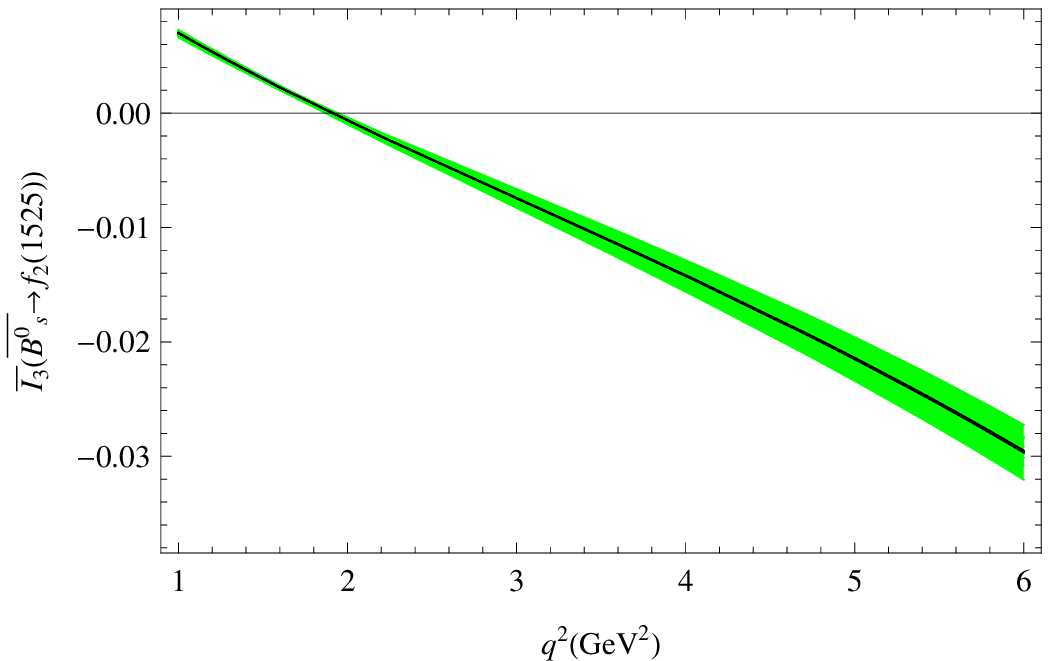} 
\hspace{1.cm}
\includegraphics[scale=0.7]{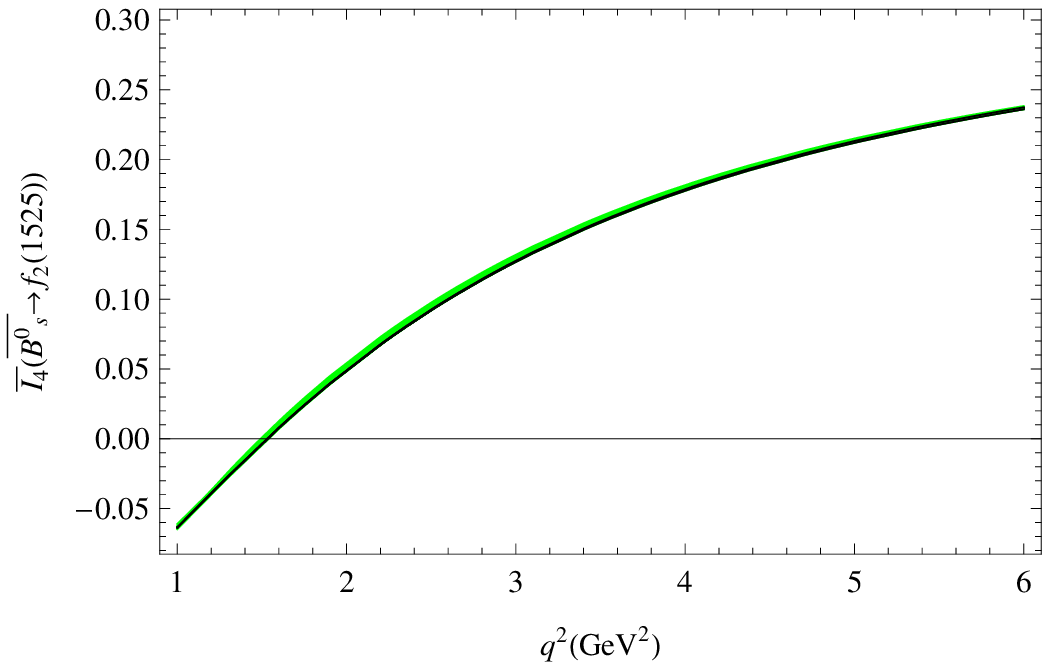} \\
\includegraphics[scale=0.7]{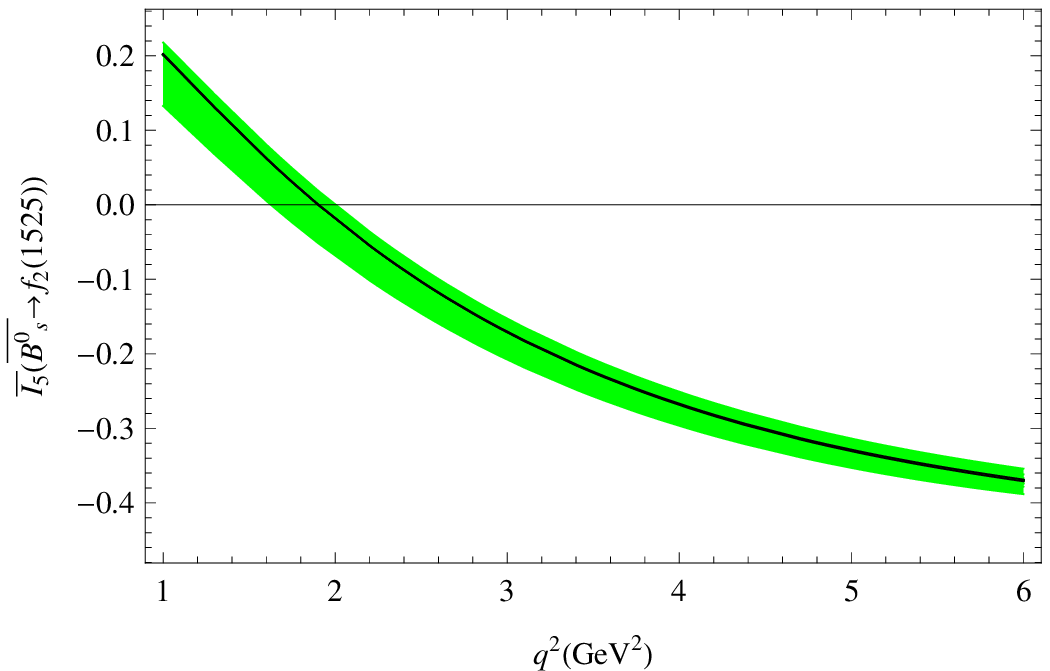} 
\hspace{1.cm}
\includegraphics[scale=0.7]{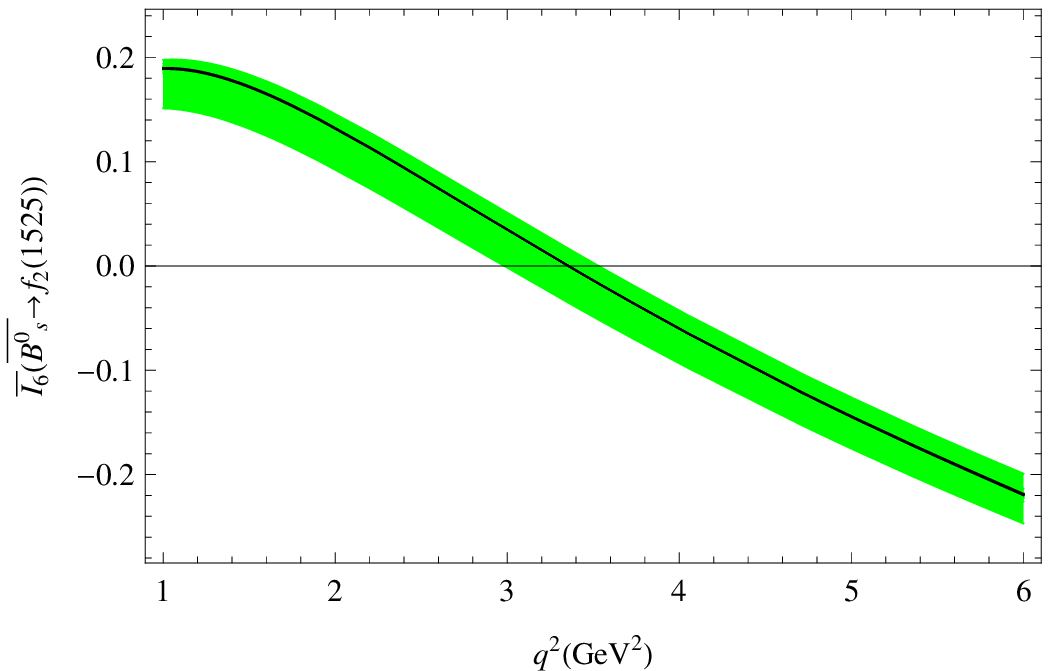} \\
\caption{Similar as Fig.~\ref{Fig:angular-coefficients-B-K2-mumu}
but for $B_s\to
f_2\mu^+\mu^-$}\label{Fig:angular-coefficients-Bs-f2-mumu}
\end{center}
\end{figure}

\begin{figure}\begin{center}
\includegraphics[scale=0.7]{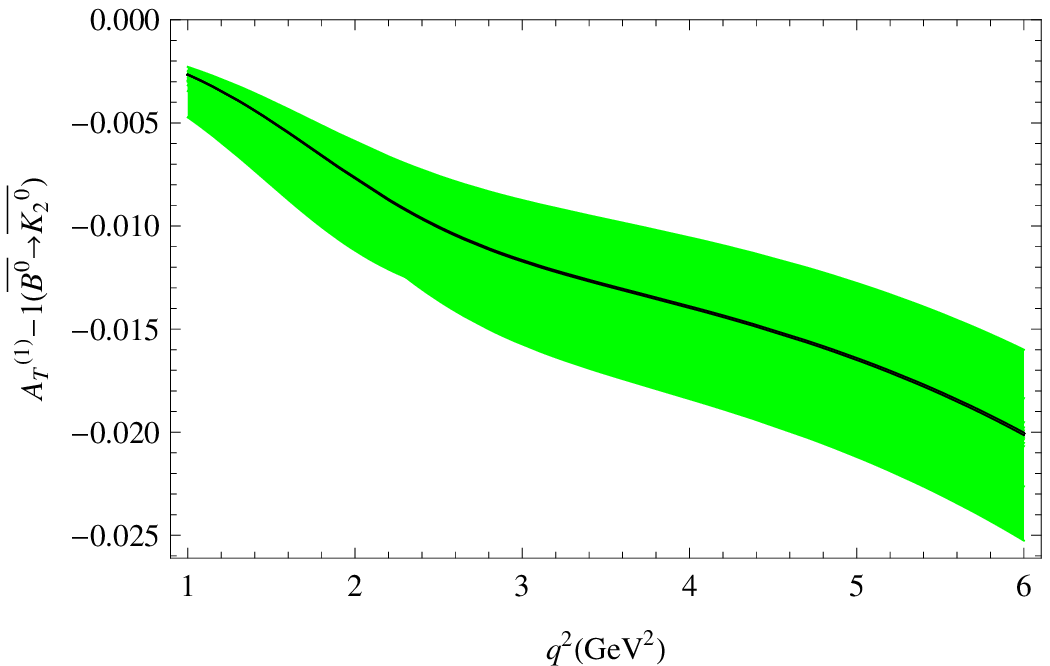} 
\hspace{1.cm}
\includegraphics[scale=0.7]{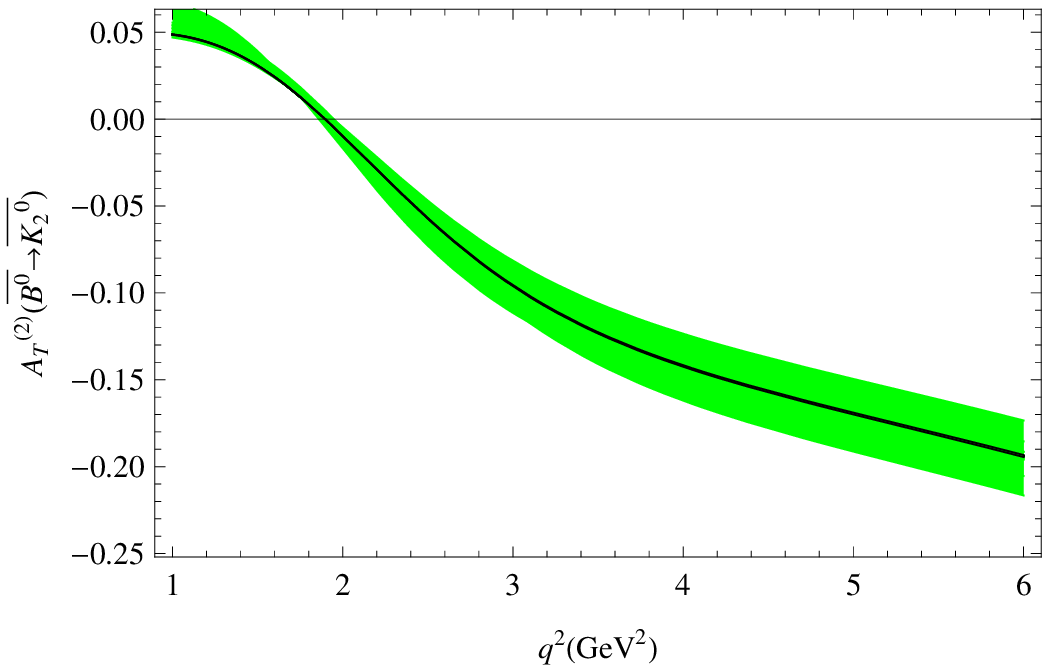} \\
\includegraphics[scale=0.7]{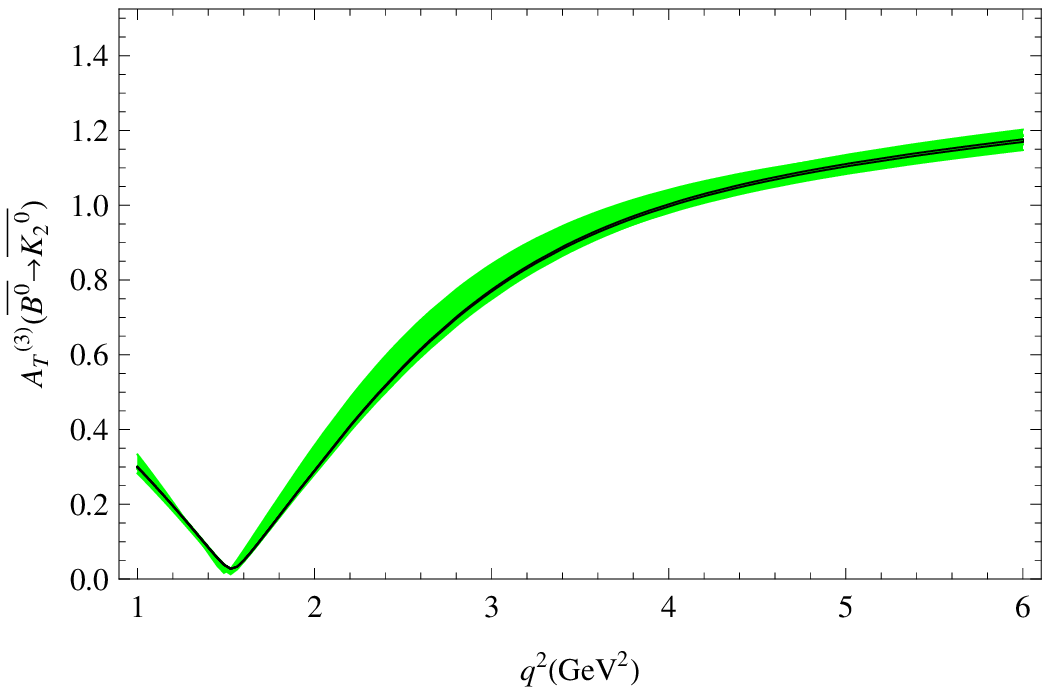} 
\hspace{1.cm}
\includegraphics[scale=0.7]{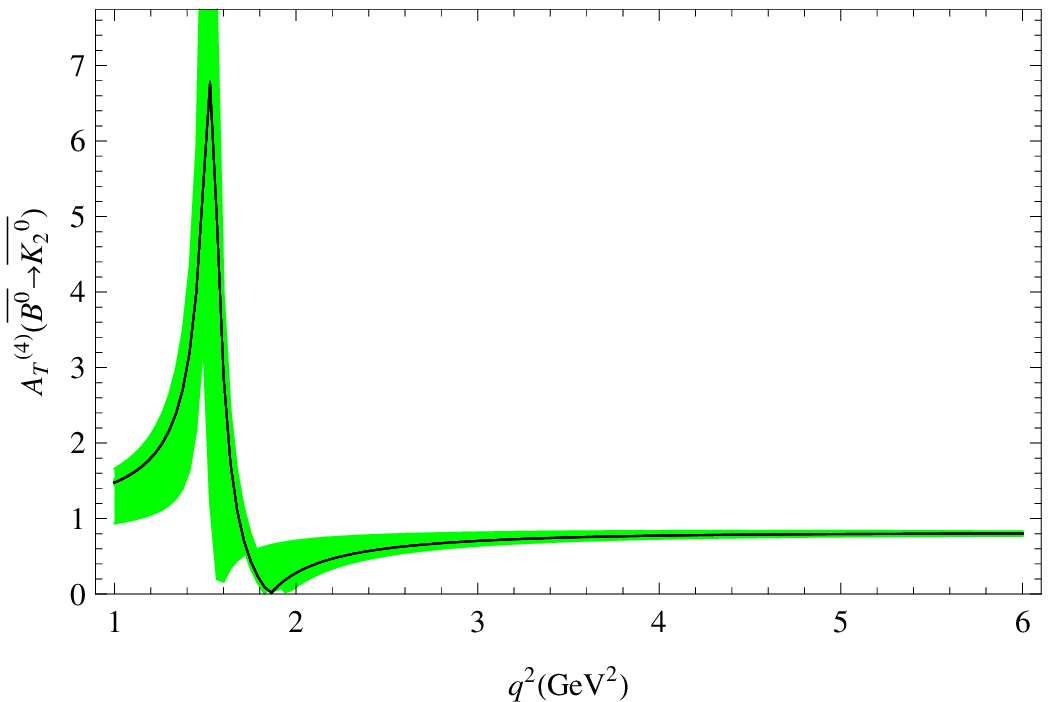} \\
\caption{Spin amplitudes and transversity asymmetries of $B\to
K_2^*\mu^+\mu^-$}\label{Fig:transvere-asymmetry-B-K2-mumu}
\end{center}
\end{figure}

\begin{figure}\begin{center}
\includegraphics[scale=0.7]{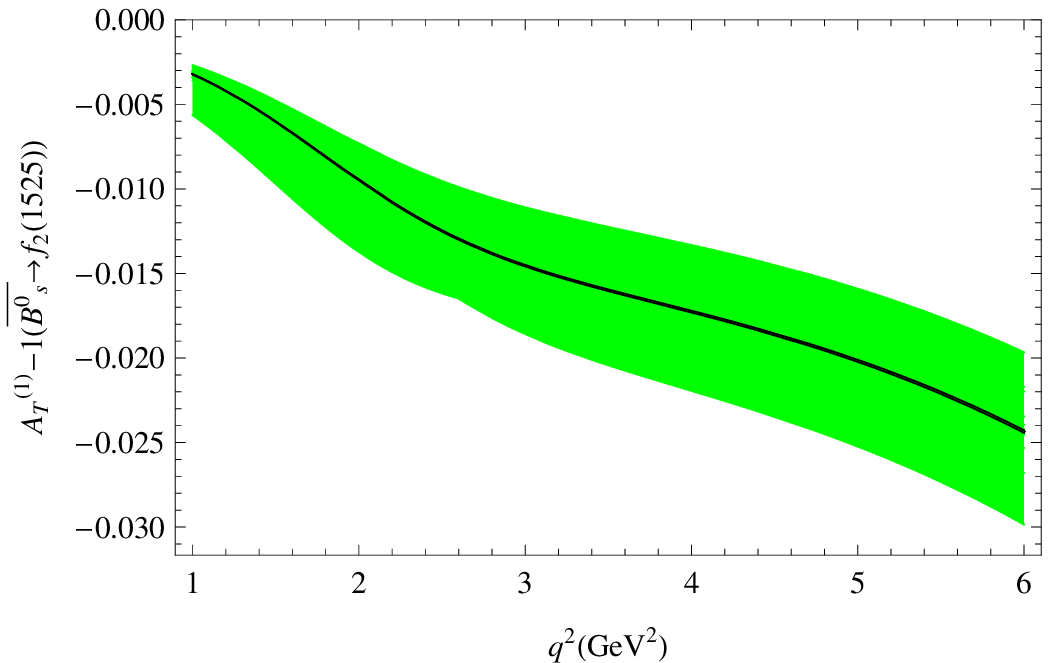} 
\hspace{1.cm}
\includegraphics[scale=0.7]{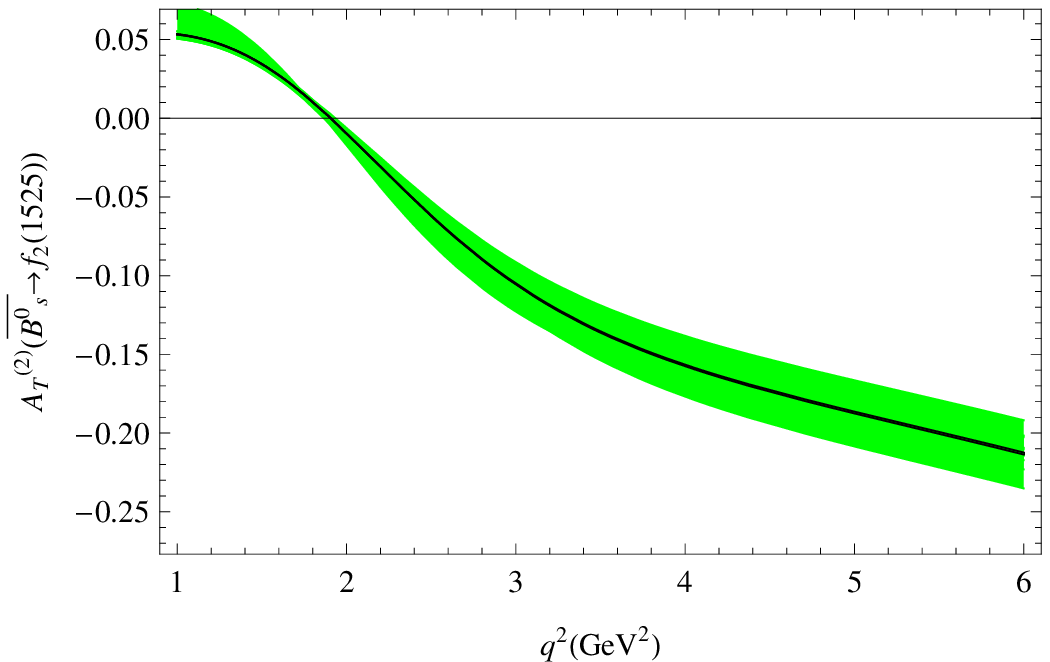} \\
\includegraphics[scale=0.7]{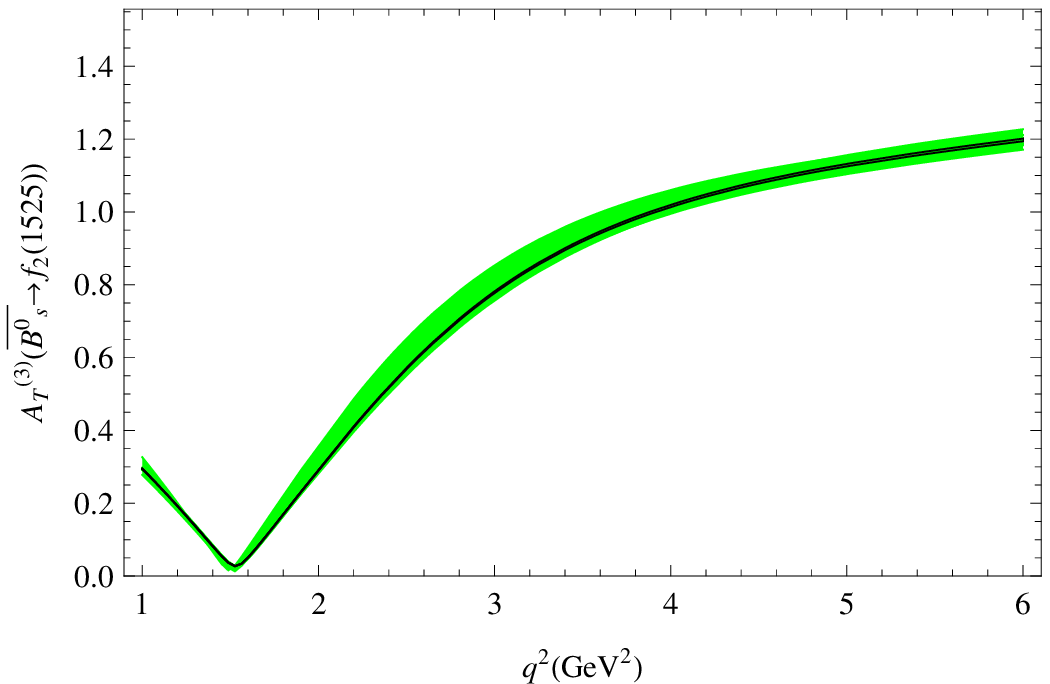} 
\hspace{1.cm}
\includegraphics[scale=0.7]{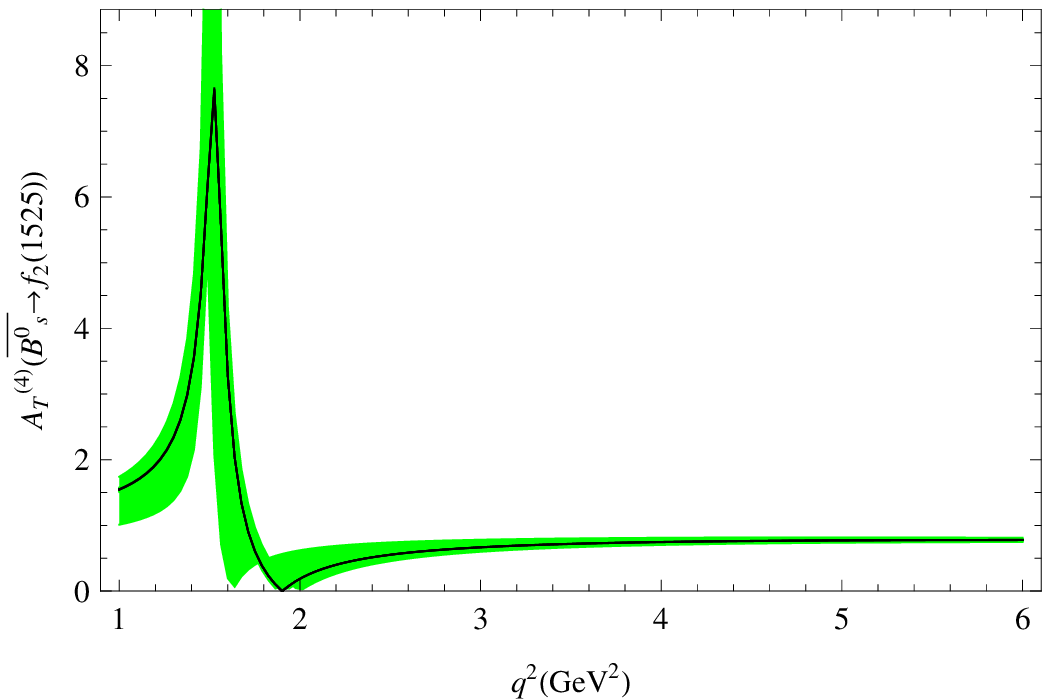} \\
\caption{Similar with Fig.~\ref{Fig:transvere-asymmetry-B-K2-mumu}
but for $B_s\to
f_2\mu^+\mu^-$}\label{Fig:transvere-asymmetry-Bs-f2-mumu}
\end{center}
\end{figure}

Our results for angular coefficients, $\bar
I_i=I_i/\frac{d\Gamma}{dq^2}$, are depicted in
Fig.~\ref{Fig:angular-coefficients-B-K2-mumu} for $B\to
K_2^*\mu^+\mu^-$ and Fig.~\ref{Fig:angular-coefficients-Bs-f2-mumu}
for $B_s\to f_2'\mu^+\mu^-$. Since the predictions for $\bar
I_7,\bar I_8,\bar I_9$ in the SM are typically smaller than 0.03, we shall not
show them.  The corresponding transversity asymmetries are shown in
Fig.~\ref{Fig:transvere-asymmetry-B-K2-mumu} and
Fig.~\ref{Fig:transvere-asymmetry-Bs-f2-mumu} respectively. One
particular feature is that most of these results are stable against
the large uncertainties from the form factors.

For the experimental purpose, it is valuable to estimate the minimum
size of the averaged value of an angular distribution coefficient
so that it can be measured in experiment. To establish any generic asymmetry
with the averaged value $\langle A\rangle$ of a particular decay at
$n\sigma$ level, events of the number $N=n^2/(\langle
A\rangle)^2$ should be accumulated. For instance on the LHCb there
are 6200 events for the $B\to K^*l^+l^-$ process per nominal running
year~\cite{:2009ny}. Incorporating all differences between $K_2^*$
and $K^*$, we may expect  roughly1000 events of   $B\to K_2^*(\to K\pi)l^+l^-$. Therefore if
one wants to observe an asymmetry  at $n\sigma$
level, its averaged value should be larger than $\langle
A\rangle_{\rm min}=\sqrt{\frac{n^2}{ 1000}}\simeq 0.03n$. 

Before closing this subsection, it is necessary to point out that the above estimation might be too optimistic.
In the first few running years of LHCb, the central energy in the $pp$ collision may not reach 14 TeV and its luminosity will be below $2 fb^{-1}$. 
Thus in the first stage not enough data are available for a precise determination of some angular coefficients.  Nevertheless this will not affect  our analysis  of branching fractions and many angular coefficients.

\subsection{Constraint on NP parameters and the NP effects on $B\to K_2^*\mu^+\mu^-$}

In this subsection we will first update constraints of free
parameters in the above two NP models, and particularly we use the
experimental data of $b\to sl^+l^-$ and $B\to K^*l^+l^-$. Decay
width of the inclusive process $b\to sl^+l^-$ is given
as~\cite{Buchalla:1995vs}
\begin{eqnarray}
 \frac{d\Gamma(b\to s \ell^+ \ell^-)}{d\hat s}&=& \Gamma(b\to c e\bar \nu_e)
 \frac{|V^*_{ts} |^2}{|V_{cb}|^2}\frac{\alpha_{\rm em}^2}{4\pi^2} \frac{(1-\hat s)^2}{f(\hat m_c) k(\hat m_c)}
 \nonumber \\
 &\times& \left[ (1+2\hat s) \left(|C_9|^2 + |C_{10}|^2 \right)
 +4\left( 1+\frac{2}{\hat s}\right) |C_{7}|^2 + 12 C_7 {\rm  Re}C_9\right]\,, \nonumber \\
 f(z)&=&1-8z^2+8z^6 -z^8 -24 z^4 \ln z\,, \nonumber \\
 k(z)&=& 1-\frac{2\alpha_s}{3\pi} \left[\left(\pi^2 -
\frac{31}{4}\right) (1-z)^2 + \frac{3}{2} \right]\,,
 \end{eqnarray}
where $\hat s=q^2/m^2_b$, and $\Gamma(b\to c e \bar \nu _e)$ is used
to cancel the uncertainties from the CKM matrix elements and the
factor $m^5_b$. For $B\to K^*l^+l^-$,  the FBAs, polarizations,
and BR  have been measured in different
kinematic bins~\cite{:2009zv} .  The other relevant experimental data collected in
Tab.~\ref{tab:inputs-data} are from Refs.~\cite{Amsler:2008zz,HFAG}.

\begin{table}[httb]
\caption{Experimental data used in the least $\chi^2$-fitting method
} \label{tab:inputs-data}
\begin{ruledtabular}
\begin{tabular}{ccccc}
 &$b\to cl\bar\nu$~\cite{Amsler:2008zz} &$b\to sl^+l^-$~\cite{HFAG}  & $\overline B^0\to K^*l^+l^-$~\cite{HFAG} \\
  &$(10.58\pm0.15)\times 10^{-2}$ & $(3.66^{+0.76}_{-0.77})\times 10^{-6}$ &$(1.09^{+0.12}_{-0.11})\times
  10^{-6}$ &\\\hline
  &$q^2 ({\rm GeV}^2)$ & ${\cal B}(10^{-7})$ & $F_L$ &$-A_{FB}$~\footnote{The different convention on $\theta_l$
 introduces a minus sign to the forward-backward asymmetry.}\\\hline
 &$[0,2]$        & $1.46\pm0.41$ & $0.29\pm0.21$ & $0.47\pm0.32$\\
 &$[2,4.3]$      & $0.86\pm0.32$ & $0.71\pm0.25$ & $0.11\pm0.37$\\
 &$[4.3,8.68]$   & $1.37\pm0.61$ & $0.64\pm0.25$ & $0.45\pm0.26$\\
 &$[10.09,12.86]$& $2.24\pm0.48$ & $0.17\pm0.17$ & $0.43\pm0.20$\\
 &$[14.18,16]$   & $1.05\pm0.30$ &$-0.15\pm0.28$ & $0.70\pm0.24$\\
 &$>16$          & $2.04\pm0.31$ & $0.12\pm0.15$ & $0.66\pm0.16$\\
 &$[1,6]$        & $1.49\pm0.47$ & $0.67\pm0.24$ & $0.26\pm0.31$
\end{tabular}
\end{ruledtabular}
\end{table}

We will adopt a least-$\chi^2$ fitting method to constrain the free
parameters, in which  the $\chi^2$ is
defined by
\begin{eqnarray}
 \chi^2_i=\frac{(B_{i}^{the}-B_{i}^{exp})^2}{(B_{i}^{err})^2},
\end{eqnarray}
where $B_i$ denotes one generic quantity among the physical
observables. The $B_{i}^{the}$,$B_{i}^{exp}$  and $B_{i}^{err}$
denotes the theoretical prediction, the cental value and 1-$\sigma$
error of experimental data, respectively. The total $\chi^2$ is
obtained by adding the individual ones. It is necessary to point out
that although the errors in experiment may correlate, for instance
the measurement of ${\cal B}$, $f_L$ and $A_{FB}$ proceed at the
same time in the fitting of angular distributions~\cite{:2009zv}, we
have not taken into account their correlation in our theoretical results.

As shown in the previous section, these two NP models have the
similarity that only $C_{9,10}$ are modified. One difference
lies in the coupling with the leptons, the newly introduced
down-type quark in VQM  will not modify the lepton sector and the
coupling with leptons is SM-like; on the contrary, one new gauge
boson is added in the $Z'$ model and its coupling with leptons  are
completely unknown.

Embedded in the VQM, the two parameters, real part and imaginary
part of $\lambda_{sb}$, are found as
\begin{eqnarray}
 {\rm Re}\lambda_{sb}=(0.07\pm0.04)\times 10^{-3},\;\;\;
 {\rm Im}\lambda_{sb}=(0.09\pm0.23)\times 10^{-3},\;\;\;
\end{eqnarray}
from which we obtain $|\lambda_{sb}|<0.3\times 10^{-3}$ but the
phase is less constrained again. The corresponding constraint on
Wilson coefficients are
\begin{eqnarray}
 |\Delta C_9| =|C_9-C_9^{SM}|<0.2,\;\;\;
 |\Delta C_{10}| =|C_{10}-C_{10}^{SM}|<2.8.
\end{eqnarray}
Our result of $\chi^2/d.o.f.$ in the fitting
method is $49.3/(23-2)$.

Turning to family nonuniveral $Z'$ model in which the coupling
between $Z'$ and a lepton pair is unknown, the two Wilson
coefficients, $C_9$ and $C_{10}$, can be chosen as independent
parameters. Assuming
$\Delta C_9$ and $\Delta C_{10}$ as real, we find 
\begin{eqnarray}
 \Delta C_9=0.88\pm0.75,\;\;\; \Delta C_{10}=0.01\pm0.69,
\end{eqnarray}
with $\chi^2/d.o.f.=48.4/(23-2)$. Removal of the above assumption
leads to
\begin{eqnarray}
 \Delta C_9=-0.81\pm1.22+(3.05\pm0.92)i,\;\;\; \Delta
 C_{10}=1.00\pm1.28+(-3.16\pm0.94)i
\end{eqnarray}
with $\chi^2/d.o.f.=45.6/(23-4)$.  If the $\mu$-lepton mass is
neglected, the imaginary part of $C_{10}$ will not appear in the
expressions for the differential decay widths and the polarizations.
Moreover, for the forward-backward asymmetry as shown in
Eq.~\eqref{eq:AFB-simplication-massless}, the imaginary part of
$C_{10}$ contributes in the combination ${\rm Re}[C_9C_{10}]$,  thus
the inclusion of ${\rm Im} [C_{10}]$ will have little effect on the $\chi^2$.

\begin{figure}\begin{center}
\includegraphics[scale=0.8]{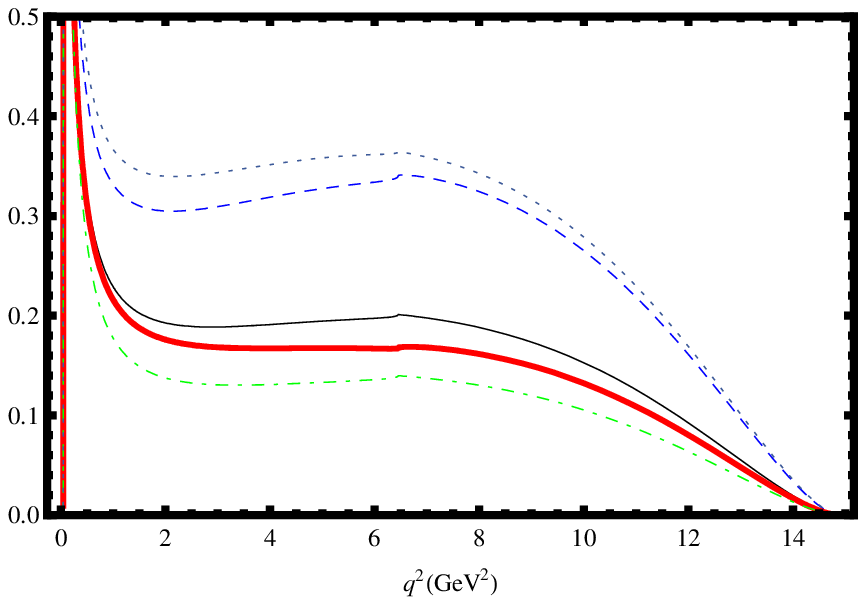} 
\hspace{1.cm}
\includegraphics[scale=0.8]{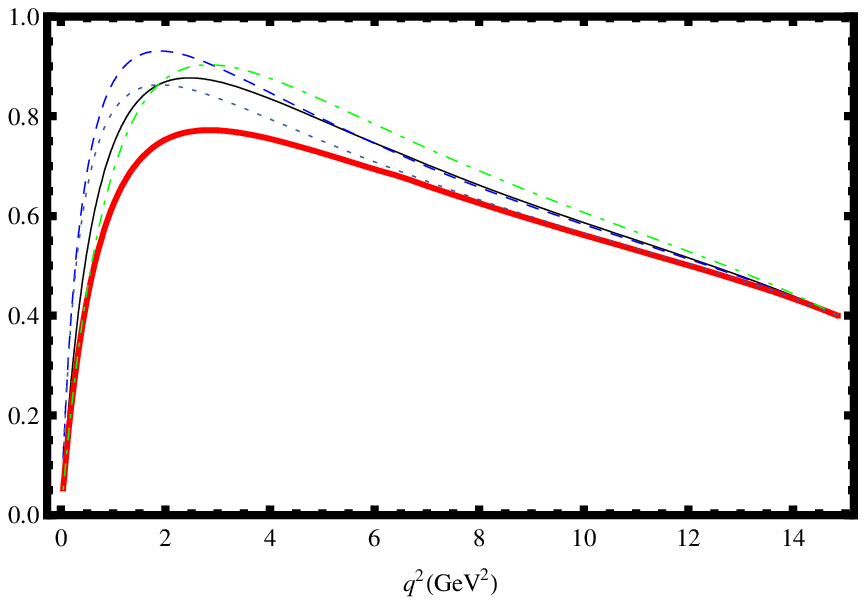} \\
\vspace{1.cm}
\includegraphics[scale=0.8]{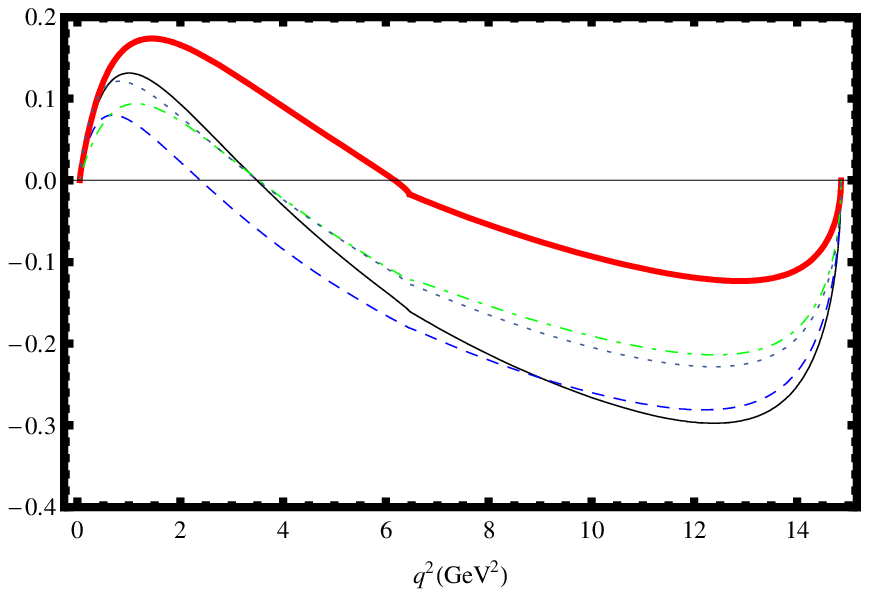} 
\caption{The impacts of the NP contributions on differential
branching ratios (in unit of $10^{-7}$), polarization fractions and
normalized forward-backward asymmetry of $B\to
K_2^*l^+l^-$}\label{Fig:B-K2-NP-C9-C10}
\end{center}
\end{figure}

Combing the above results, we can see that the NP contributions in
both cases satisfy
\begin{eqnarray}
 |\Delta C_9|<3,\;\;\;
 |\Delta C_{10}|<3.
\end{eqnarray}
To illustrate, we choose
$\Delta C_9=3e^{i\pi/4,i3\pi/4}$ and $\Delta
C_{10}=3e^{i\pi/4,i3\pi/4}$ as the reference points and give the
plots of branching ratios, FBAs and the polarizations in
Fig.~\ref{Fig:B-K2-NP-C9-C10}. The black (solid) line denotes the SM
result, while the dashed (blue) and thick (red) lines correspond to
the modification of $C_9$. The dot-dashed (green) and dotted lines
are obtained by modifying $C_{10}$. From the figure for $A_{FB}$, we can see that
the zero-crossing point  $s_0$ can be sizably changed, which can be
tested on the future collider or can be further constrained.

One last process to explore is $B_s\to \mu^+\mu^-$, of which the
branching fraction is
\begin{eqnarray}
 {\cal B}( B_s \to \mu^+ \mu^-) &=& \tau_{B_s}\frac{G^2_F\alpha^2_{\rm em}}{16\pi^3} |V^*_{ts}  V_{tb}|^2
 m_{B_s}f^2_{B_s}m^2_{\mu}|C_{10}|^2 \left(1-\frac{4m^2_{\mu}}{m^2_{B_s}}\right)^{1/2}.
\end{eqnarray}
Using the same inputs as those in our computation of $B\to
K_2^*l^+l^-$, we have
\begin{eqnarray}
 {\cal B}( B_s \to \mu^+ \mu^-) &=&3.50\times 10^{-9} \left(\frac{f_{B_s}}{230 {\rm MeV}}\right)^2
 \left(\frac{|C_{10}|}{4.67}\right)^2.
\end{eqnarray}
Even if $C_{10}$ is enhanced by a factor of 2, the above result is
still consistent with the recent measurement~\cite{Abazov:2010fs}
\begin{eqnarray}
 {\cal B}( B_s \to \mu^+ \mu^-) <5.1\times 10^{-8}.
\end{eqnarray}

\section{Summary}

In this work we have explored $B\to K_2^*(\to K\pi)l^+l^-$ (with
$l=e,\mu,\tau$) decays and a similar mode $B_s\to f_2'(1525)(\to
K^+K^-) l^+l^-$ in the standard model and two new physics scenarios:
vector-like quark model and family non-universal $Z'$ model.  Besides
branching ratios, forward-backward asymmetries and transversity
amplitudes, we have also derived the differential angular
distributions of this decay chain. The sizable production rates  lead to a promising
prospective to observe this channel on the future experiment.

Using the experimental data of the inclusive $b\to sl^+l^-$ and
$B\to K^*l^+l^-$, we have updated the constraints on effective
Wilson coefficients and/or free parameters in these two new physics
scenarios. In the VQM, we find that the constraint on the coupling
constant is improved by a factor of 3 compared with our previous
work. Their impact on $B\to K_2^*l^+l^-$ is elaborated and in
particular the zero-crossing point for the forward-backward asymmetry in
these NP scenarios can sizably deviate from the SM. These results
will be tested on the future hadron collider.

\section*{Acknowledgements}

This work  is partly  supported by National Natural Science Foundation of
China under Grant Nos. 10735080,
11075168 and 10625525; and National Basic Research Program of China
(973) No. 2010CB833000. (C.D. L\"u), the
Brain Korea 21 Project (R.H. Li) and Alexander von Humboldt
Stiftung (W. Wang). We thank K.C. Yang for useful discussion on the coefficient ${\cal C}$ defined in the appendix.  W. Wang is grateful to Prof. Ahmed Ali for
valuable discussions, C.H. Chen for the collaboration and useful discussions in the
vector-like quark model, and Yuming Wang for useful discussions
of charm-loop effects.

\begin{appendix}

\section{Effective Hamiltonian}

The effective Hamiltonian governing $b\to sl^+l^-$ is given by
 \begin{eqnarray}
 {\cal
 H}_{\rm{eff}}=-\frac{G_F}{\sqrt{2}}V_{tb}V^*_{ts}\sum_{i=1}^{10}C_i(\mu)O_i(\mu),\label{eq:Hamiltonian}
 \end{eqnarray}
where $V_{tb}=0.999176$ and $V_{ts}=-0.03972$~\cite{Amsler:2008zz}
are the CKM matrix elements and $C_i(\mu)$ are Wilson
coefficients for the effective operators $O_i$. In this paper, we
will adopt the Wilson coefficients up to the leading
logarithmic accuracy~\cite{Buchalla:1995vs}, and their values in SM are
listed in Tab.~\ref{tab:wilsons}. Since the NP scenarios considered
in the present paper would not introduce any new operator, the SM operators
will form a complete basis for our analysis
 \begin{eqnarray}
 O_1&=&(\bar s_{\alpha}c_{\alpha})_{V-A}(\bar
 c_{\beta}b_{\beta})_{V-A},\;\;
 O_2=(\bar
 s_{\alpha}c_{\beta})_{V-A}(\bar
 c_{\beta}b_{\alpha})_{V-A},\nonumber\\
 O_3&=&(\bar s_{\alpha}b_{\alpha})_{V-A}\sum_q(\bar
 q_{\beta}q_{\beta})_{V-A},\;\;
 O_4=(\bar s_{\alpha}b_{\beta})_{V-A}\sum_q(\bar
 q_{\beta}q_{\alpha})_{V-A},\nonumber\\
 O_5&=&(\bar s_{\alpha}b_{\alpha})_{V-A}\sum_q(\bar
 q_{\beta}q_{\beta})_{V+A},\;\;
 O_6=(\bar s_{\alpha}b_{\beta})_{V-A}\sum_q(\bar
 q_{\beta}q_{\alpha})_{V+A},\nonumber\\
 O_7&=&\frac{e m_b}{8\pi^2}\bar
 s\sigma^{\mu\nu}(1+\gamma_5)bF_{\mu\nu}+\frac{e m_s}{8\pi^2}\bar
 s\sigma^{\mu\nu}(1-\gamma_5)bF_{\mu\nu},\nonumber\\
 O_9&=&\frac{\alpha_{\rm{em}}}{2\pi}(\bar l\gamma_{\mu}l)(\bar
 s\gamma^{\mu}(1-\gamma_5)b),\;\;
 O_{10}=\frac{\alpha_{\rm{em}}}{2\pi}(\bar l\gamma_{\mu}\gamma_5l)(\bar
 s\gamma^{\mu}(1-\gamma_5)b).\label{eq:operators}
 \end{eqnarray}
The left-handed and right-handed operators are $(\bar
q_1q_2)_{V-A}(\bar q_3 q_4)_{V\pm A}\equiv(\bar q_1
\gamma^{\mu}(1-\gamma_5)q_2)(\bar
q_3\gamma_{\mu}(1\pm\gamma_5)q_4)$. $m_b=4.8$GeV  and $m_s=0.095$GeV are $b$ and $s$ quark masses in the $\overline{\mbox{MS}}$ scheme and 
 $\alpha_{\rm em}=1/137$ is fine structure
constant. The double
Cabibbo suppressed terms, proportional to $V_{ub}V_{us}^*$, have
been neglected.

 \begin{table}
 \caption{The values of Wilson coefficients $C_i(m_b)$ in the leading
logarithmic approximation, with $m_W=80.4\mbox{GeV}$, $\mu=m_{b,\rm
pole}$~\cite{Buchalla:1995vs}.}
 \label{tab:wilsons}
 \begin{center}
 \begin{tabular}{c c c c c c c c c}
 \hline\hline
 \ \ \ $C_1$ &$C_2$ &$C_3$ &$C_4$ &$C_5$ &$C_6$ &$C_7^{\rm{eff}}$ &$C_9$ &$C_{10}$       \\
 \ \ \ $1.107$   &$-0.248$   &$-0.011$    &$-0.026$    &$-0.007$    &$-0.031$    &$-0.313$    &$4.344$    &$-4.669$    \\
 \hline\hline
 \end{tabular}
 \end{center}
 \end{table}

At the one-loop level accuracy, the matrix element of $b\to sl^+l^-$
transition receives loop contributions from $O_1-O_6$. Since the
factorizable loop terms~\cite{Buras:1994dj} can be incorporated into
the Wilson coefficients $C_7$ and $C_{9}$, it is convenient to
define  combinations $C_7^{\rm{eff}}$ and
$C_9^{\rm{eff}}$~\cite{Buras:1994dj}
\begin{eqnarray}
 C_7^{\rm{eff}}&=&C_7-C_5/3-C_6,\nonumber\\
 C_9^{\rm{eff}}(q^2)&=&C_9(\mu)+h(\hat{m_c},\hat{s})C_0-\frac{1}{2}h(1,\hat{s})(4C_3+4
 C_4+3C_5+C_6)\nonumber\\
 &&-\frac{1}{2}h(0,\hat{s})(C_3+3
 C_4) + \frac{2}{9}(3C_3 + C_4 +3C_5+ C_6),\label{eq:C7C9eff}
\end{eqnarray}
with $\hat{s}=q^2/m_b^2$, $C_0=C_1+3C_2+3C_3+C_4+3C_5 +C_6$, and
$\hat{m}_c=m_c/m_b$.  The auxiliary functions used  above are
\begin{eqnarray}
 h(z,\hat s)&=& -\frac{8}{9}\ln \frac{m_b}{\mu}-\frac{8}{9}\ln
 z+\frac{8}{27}+\frac{4}{9}x-\frac{2}{9}(2+x)|1-x|^{1/2}
 \left\{\begin{array}{c}
 \ln\left| \frac{\sqrt{1-x}+1}{\sqrt{1-x}-1}\right|-i\pi \;\;\; {\rm for} \;\; x\equiv \frac{4z^2}{\hat s}<1 \\
  2{\rm arctan}\frac{1}{\sqrt{x-1}}\;\;\; {\rm for} \;\; x\equiv \frac{4z^2}{\hat s}>1
 \end{array}\right.,\nonumber\\
 h(0,\hat s)&=& -\frac{8}{9}\ln \frac{m_b}{\mu}-\frac{4}{9}\ln
 \hat s+\frac{8}{27}+\frac{4}{9}i\pi.
\end{eqnarray}
In the following, we shall also drop the superscripts for $C_{9}^{\rm
eff}$ and $C_{7}^{\rm eff}$ for convenience.


On the hadron level resonant states, such as vector charmonia
generated from the $b \to c\bar cs$, may annihilate into a lepton
pair. Therefore  they will also contribute in a long distance
manner~\cite{Lim:1988yu,Ali:1991is,Lu_and_zhang}. But these
contributions can be subtracted with a kinematic cutoff in
experiment. Moreover our following analysis of differential
distributions will be mainly dedicated to the region of $1{\rm
GeV}^2< q^2<6{\rm GeV}^2$, also excluding contributions from the
charmonia.


\section{Helicity amplitudes}

Within a graphic picture $B\to K_2^*(\to K\pi)l^+l^-$ proceeds
via three steps: $B$ meson first decays into an onshell strange
meson plus a pair of leptons; the $K_2^*$ meson  propagates followed
by its strong decay into $K\pi$. To evaluate the decay
width of multibody decays, we shall adopt the helicity amplitude
which mainly uses
\begin{eqnarray}
 g_{\mu\nu}=-\sum_\lambda\epsilon_\mu(\lambda) \epsilon^*_\nu(\lambda) +\frac{q_\mu q_\nu
 }{q^2}.\label{eq:replacement}
\end{eqnarray}
$\epsilon$ is the polarization vector with the momentum $q$ and
$\lambda$ denotes the three kinds of polarizations. The last term
can be formally identified as a timelike polarization
$\epsilon_\mu(t)=\frac{q_\mu}{\sqrt {q^2}}$,  and thus the metric tensor
$g_{\mu\nu}$ can be then understood as summations of the four
polarizations. For the purpose of illustration we will first
evaluate the decay amplitude of $B\to K_2^*l^+l^-$. In the SM, the
lepton pair in the final state is produced via an off-shell photon,
a $Z$ boson or some hadronic vector mesons. These states may have
different couplings but they share many commonalities: the Lorentz
structure for the vertex of the lepton pair is either $V-A$ or $V+A$
or some combination of them. Therefore the decay amplitudes of $\bar
B\to \bar K_2^*l^+l^-$ can be rewritten as
\begin{eqnarray}
 {\cal A}(\bar B\to \bar K_2^*l^+l^-)&=& {\cal L}^\mu(L) {\cal H}_\mu(L)+{\cal L}^\mu(R) {\cal H}_\mu(R),
\end{eqnarray}
in which ${\cal L}_\mu(L),{\cal L}_\mu(R)$ are the lepton pair spinor products:
\begin{eqnarray}
 {\cal L}_\mu(L)&=& \bar l\gamma_\mu(1-\gamma_5)l,\;\;\;
 {\cal L}_\mu(R)= \bar l\gamma_\mu(1+\gamma_5)l,
\end{eqnarray}
while ${\cal H}$ incorporates the remaining  $B\to K_2^*$ part. In
the case of massless leptons, left-handed and right-handed
sectors decouple, which will greatly simplify the analysis.   The
identity in Eq.~\eqref{eq:replacement} results in a factorization of
decay amplitudes
\begin{eqnarray}
 {\cal A}(\bar B\to \bar K_2^*l^+l^-)&=& {\cal L}_\mu(L) {\cal H}_\nu(L) g^{\mu\nu}+{\cal L}_\mu(R) {\cal H}_\nu(R)
 g^{\mu\nu}\nonumber\\
  &=&-\sum_{\lambda}{\cal L}_{L\lambda} {\cal H}_{L\lambda}  -\sum_{\lambda}{\cal L}_{R\lambda} {\cal
  H}_{R\lambda},
\end{eqnarray}
where $q^\mu$ is the momentum of the lepton pair and ${\cal
L}_{L\lambda}={\cal L}^\mu(L) \epsilon_\mu(\lambda)$ and ${\cal
L}_{R\lambda}={\cal L}^\mu(R) \epsilon_\mu(\lambda)$ denote
Lorentz invariant amplitudes for the lepton part. It is also similar
for the Lorentz invariant hadronic amplitudes: ${\cal
H}_{L\lambda}={\cal H}^\mu(L) \epsilon^*_\mu(\lambda)$ and ${\cal
H}_{R\lambda}={\cal H}^\mu(R) \epsilon^*_\mu(\lambda)$. The timelike
polarization gives vanishing contributions in the case of $m_l=0$
for $l=e,\mu$: using equation of motion, this term is proportional
to the lepton mass.


An advantage of the helicity amplitudes is that both hadronic
amplitudes and leptonic amplitudes are Lorentz invariant. Such a
good property allows  to choose different frames in the evaluation.
For instance leptonic amplitudes are evaluated in the lepton pair
central mass frame, while hadronic $B$ decay amplitudes are directly
obtained in the $B$ rest frame.  Since $K_2^*$ and $K^*$ have several
important similarities, $B\to K_2^*l^+l^-$ differential decay widths
can be simply obtained from the ones of $B\to K^*l^+l^-$ in a comparative manner.
\begin{itemize}
\item
Longitudinal and transverse $B$ decay amplitudes are obtained by
multiplying the factor $\frac{\sqrt{\lambda}}{\sqrt8 m_Bm_{K_2^*}}$
and $\frac{\sqrt{\lambda}}{\sqrt6m_Bm_{K_2^*}}$ respectively. The
function $\lambda$ is the magnitude of the $K_2^*$ momentum in
$B$ meson rest frame: $\lambda\equiv\lambda(m^2_{B},m^2_{K_2^*},
q^2)=2m_B|\vec p_{K_2^*}|$, and
$\lambda(a^2,b^2,c^2)=(a^2-b^2-c^2)^2-4b^2c^2$. This replacement is an
output of the fact that the polarization vector $\epsilon$ is
replaced by $\epsilon_T$ in the form factor definitions. Explicitly, these hadronic amplitudes are 
\begin{eqnarray}
 H_{L0}
  &=& N  \frac{\sqrt{\lambda}}{\sqrt8 m_Bm_{K_2^*}}\frac{1}{2m_{K^*_2}\sqrt {q^2}}\left[ (C_9-C_{10})
[(m_B^2-m_{K^*_2}^2-q^2)(m_B+m_{K^*_2})A_1
 -\frac{\lambda}{m_B+m_{K^*_2}}A_2]\right.\nonumber\\
 &&\left. +  2m_b(C_{7L}-C_{7R})  [ (m_B^2+3m_{K_2^*}^2-q^2)T_2 -\frac{\lambda  }
 {m_B^2-m_{K_2^*}^2}T_3]\right],\nonumber\\
 H_{L\pm}
 &=& N \frac{\sqrt{\lambda}}{\sqrt6m_Bm_{K_2^*}}
  \left[ (C_9-C_{10}) [(m_B+m_{K^*_2})A_1\mp \frac{\sqrt \lambda}{m_B+m_{K^*_2}}V]\right.\nonumber\\
 &&\left.
 -\frac{2m_b(C_{7L}+C_{7R})}{q^2} (\pm\sqrt \lambda T_1)+\frac{2m_b(C_{7L}-C_{7R})}{q^2}
 (m_B^2-m_{K_2^*}^2)T_2\right],
 \nonumber\\
 H_{Lt}&=& N   \frac{\sqrt{\lambda}}{\sqrt8 m_Bm_T}
 (C_{9}-C_{10})\frac{\sqrt \lambda}{\sqrt {q^2}}A_0, \nonumber\\
 H_{Ri}
  &=& H_{Li}|_{C_{10}\to -C_{10}}
\end{eqnarray}
with $N=-i\frac{G_F} {4\sqrt 2}\frac{\alpha_{\rm em}}{\pi}
V_{tb}V_{ts}^*$.

\item
In the propagation of the intermediate strange meson, the width effect of
$K_2^*$ could be more important since $\Gamma _{K_2^*}\sim 100 {\rm
MeV}>\Gamma_{K^*}\sim 50 {\rm MeV}$~\cite{Amsler:2008zz}.
Nevertheless, since the  $K_2^*$  width is only larger than that of $K^*$
by a factor of 2, the narrow-width approximation, which has been
well used in the case of $K^*$, might also work for $K_2^*$. In this
sense, there is no difference 
except that the ${\cal B}(K^*\to K\pi)$ is replaced by ${\cal
B}(K^*_2\to K\pi)$.

\item
Incorporation of the $K_2^*\to K\pi$ decay gives the complete
results for differential decay distribution of $B\to K_2^*(\to
K\pi)l^+l^-$.  Angular distributions of $K_2^*$ and $K^*$ strong
decays are described by spherical harmonic functions:
$Y_{1}^{i}(\theta,\phi)$ for $K^*$ and $Y_{2}^{i}(\theta,\phi)$ for
$K_2^*$. In particular we find the relations 
\begin{eqnarray}
 &&\sqrt{\frac{3}{4\pi}} \cos(\theta_{K})\equiv C(K^*)\to \sqrt{\frac{5}{16\pi}}
 (3\cos^2\theta_{K}-1)\equiv C(K_2^*),\nonumber\\
 &&
 \sqrt{\frac{3}{8\pi}} \sin(\theta_{K})\equiv S(K^*)\to
 \sqrt{\frac{15}{32\pi}}\sin(2\theta_{K})\equiv S(K_2^*).
\end{eqnarray}
\end{itemize}

Our formulas for branching fractions and forward-backward asymmetries can be shown compatible with  the ones in Ref.~\cite{Hatanaka:2009gb} through the following relations
\begin{eqnarray}
 A_{L0}&=& N_{K_2^*}  \alpha_Lm_B^3\frac{1}{2m_{K_2^*}\sqrt {q^2}}\Big(-(1-\hat m_{K_2^*}^2-\hat q^2) {\cal F}+\hat \lambda {\cal G}  +(1-\hat m_{K_2^*}^2-\hat q^2) {\cal B}- \hat\lambda  {\cal C} \Big),\\
 A_{L\perp}&=& -\frac{\sqrt{2\lambda} N_{K_2^*}\beta_T} {2m_B} ( {\cal A}-{\cal E}),\\
 A_{L||}&=& \frac{\sqrt{2\lambda} N_{K_2^*}\beta_T} {2m_B} ({\cal B}-{\cal F}),\\
 A_{t}&=& \frac{N_{K_2^*}\alpha_L }{\hat m_{K_2^*}}\frac{\sqrt\lambda}{\sqrt{q^2}}[{\cal F}-(1-\hat m_{K_2^*}^2 {\cal G}-\hat q^2{\cal H}],
\end{eqnarray}
where the coefficients ${\cal A,B,E,F,G,H}$ are defined in  Eqs. (49,50,53, 54) in Ref.~\cite{Hatanaka:2009gb} but  
the coefficient ${\cal C}$ in Eq.~(51) contains a typo and  should be read as 
\begin{eqnarray}
 {\cal C}&=& \frac{1}{1-\hat m_{K_2^*}^2} \Big[(1-\hat m_{K_2^*}) c_9^{\rm eff} (\hat s) A_2^{K_2^*}(s) +2\hat m_b c_7^{\rm eff} \big(T_3^{K_2^*}(s)+\frac{1-\hat m_{K_2^*}^2}{\hat s} T_2^{K_2^*}(s)\big)\Big].
\end{eqnarray}    
The dimensionless constants are given as
$\hat \lambda = \lambda /m_B^4$,  $\hat m_{K_2^*}= m_{K_2^*}/m_B$, $\hat m_b=m_b/m_B$  and $\hat q^2= q^2/m_B^2$.  $\alpha_L=\sqrt {2/3}$ and $\beta_T=1/\sqrt 2$.


\end{appendix}


\begin{thebibliography}{11}






\bibitem{Aubert:2008ps}
  B.~Aubert {\it et al.}  [BABAR Collaboration],
  Phys.\ Rev.\ Lett.\  {\bf 102}, 091803 (2009)
  [arXiv:0807.4119 [hep-ex]].

\bibitem{:2009zv}
  J.~T.~Wei {\it et al.}  [BELLE Collaboration],
  Phys.\ Rev.\ Lett.\  {\bf 103}, 171801 (2009)
  [arXiv:0904.0770 [hep-ex]].

\bibitem{Aaltonen:2008xf}
  T.~Aaltonen {\it et al.}  [CDF Collaboration],
  Phys.\ Rev.\  D {\bf 79}, 011104 (2009)
  [arXiv:0804.3908 [hep-ex]].


\bibitem{:2009ny}
   B. Adeva, {\it et al.}  [LHCb Collaboration],
  arXiv:0912.4179 [hep-ex];
  M. Patel and H. Skottowe, A Fisher discriminant selection for $B_d\to K^*\mu^+\mu^-$ at LHCb,
  LHCb-2009-009.




\bibitem{Gac:2010ws}
  R.~L.~Gac, [LHCb Collaboration],
  arXiv:1009.5902 [hep-ex].


\bibitem{O'Leary:2010af}
  B.~O'Leary {\it et al.} [SuperB Collaboration ],
  [arXiv:1008.1541 [hep-ex]].

\bibitem{Ali:1999mm}
  A.~Ali, P.~Ball, L.~T.~Handoko {\it et al.},
  Phys.\ Rev.\  {\bf D61}, 074024 (2000).
  [hep-ph/9910221].


\bibitem{Kim:2000dq}
  C.~S.~Kim, Y.~G.~Kim, C.~D.~Lu and T.~Morozumi,
  Phys.\ Rev.\  D {\bf 62}, 034013 (2000)
  [arXiv:hep-ph/0001151].

\bibitem{Beneke:2001at}
  M.~Beneke, T.~Feldmann, D.~Seidel,
  Nucl.\ Phys.\  {\bf B612}, 25-58 (2001).
  [hep-ph/0106067].

\bibitem{Chen:2002bq}
  C.~H.~Chen and C.~Q.~Geng,
  Nucl.\ Phys.\  B {\bf 636}, 338 (2002)
  [arXiv:hep-ph/0203003].



\bibitem{Kruger:2005ep}
  F.~Kruger and J.~Matias,
  Phys.\ Rev.\  D {\bf 71}, 094009 (2005)
  [arXiv:hep-ph/0502060].

\bibitem{Ali:2006ew}
  A.~Ali, G.~Kramer, G.~-h.~Zhu,
  Eur.\ Phys.\ J.\  {\bf C47}, 625-641 (2006).
  [hep-ph/0601034].
  
\bibitem{Bobeth:2008ij}
  C.~Bobeth, G.~Hiller and G.~Piranishvili,
  JHEP {\bf 0807}, 106 (2008)
  [arXiv:0805.2525 [hep-ph]].

\bibitem{Egede:2008uy}
  U.~Egede, T.~Hurth, J.~Matias, M.~Ramon and W.~Reece,
  JHEP {\bf 0811}, 032 (2008)
  [arXiv:0807.2589 [hep-ph]].

\bibitem{Altmannshofer:2008dz}
  W.~Altmannshofer, {\it et al.},
  JHEP {\bf 0901}, 019 (2009).
  [arXiv:0811.1214 [hep-ph]].

\bibitem{Chiang:2009dx}
  C.~W.~Chiang, R.~H.~Li and C.~D.~Lu,
  arXiv:0911.2399 [hep-ph].



\bibitem{Alok:2009tz}
  A.~K.~Alok, A.~Dighe, D.~Ghosh, D.~London, J.~Matias, M.~Nagashima and A.~Szynkman,
  JHEP {\bf 1002}, 053 (2010)
  [arXiv:0912.1382 [hep-ph]].

\bibitem{Chang:2010zy}
  Q.~Chang, X.~Q.~Li and Y.~D.~Yang,
  JHEP {\bf 1004}, 052 (2010)
  [arXiv:1002.2758 [hep-ph]].

\bibitem{Bharucha:2010bb}
  A.~Bharucha and W.~Reece,
  Eur.\ Phys.\ J.\  C {\bf 69}, 623 (2010)
  [arXiv:1002.4310 [hep-ph]].

\bibitem{Khodjamirian:2010vf}
  A.~Khodjamirian, T.~Mannel, A.~A.~Pivovarov and Y.~M.~Wang,
  JHEP {\bf 1009}, 089 (2010)
  [arXiv:1006.4945 [hep-ph]].





\bibitem{Bobeth:2010wg}
  C.~Bobeth, G.~Hiller and D.~van Dyk,
  JHEP {\bf 1007}, 098 (2010)
  [arXiv:1006.5013 [hep-ph]].
  
\bibitem{Alok:2010zd}
  A.~K.~Alok, A.~Datta, A.~Dighe, M.~Duraisamy, D.~Ghosh, D.~London and S.~U.~Sankar,
  arXiv:1008.2367 [hep-ph].
  
  

\bibitem{Rai-Choudhury:2006gv}
  S.~Rai Choudhury, {\it et al.},
  Phys.\ Rev.\  {\bf D74}, 054031 (2006)
  [hep-ph/0607289].




\bibitem{Choudhury:2009fz}
  S.~R.~Choudhury, A.~S.~Cornell and N.~Gaur,
  Phys.\ Rev.\  D {\bf 81}, 094018 (2010)
  [arXiv:0911.4783 [hep-ph]].


\bibitem{Hatanaka:2009gb}
  H.~Hatanaka and K.~C.~Yang,
  Phys.\ Rev.\  D {\bf 79}, 114008 (2009)
  [arXiv:0903.1917 [hep-ph]].

\bibitem{Hatanaka:2009sj}
  H.~Hatanaka and K.~C.~Yang,
  Eur.\ Phys.\ J.\  C {\bf 67}, 149 (2010)
  [arXiv:0907.1496 [hep-ph]].


\bibitem{Wang:2010ni}
  W.~Wang,
  Phys.\ Rev.\ D {\bf 83}, 014008 (2011)[arXiv: 1008.5326 [hep-ph]]


\bibitem{Rizzo:1985db}
  T.~G.~Rizzo,
  Phys.\ Rev.\  D {\bf 33}, 3329 (1986).

\bibitem{Branco:1986my}
  G.~C.~Branco and L.~Lavoura,
  Nucl.\ Phys.\  B {\bf 278}, 738 (1986).


\bibitem{Langacker:1988ur}
  P.~Langacker and D.~London,
  Phys.\ Rev.\  D {\bf 38}, 886 (1988).

\bibitem{Shin:1988eu}
  M.~Shin, M.~Bander and D.~Silverman,
  Phys.\ Lett.\  B {\bf 219}, 381 (1989).





\bibitem{Nir:1990yq}
  Y.~Nir and D.~J.~Silverman,
  Phys.\ Rev.\  D {\bf 42}, 1477 (1990).
\bibitem{Barenboim:1997pf}
  G.~Barenboim and F.~J.~Botella,
  Phys.\ Lett.\  B {\bf 433}, 385 (1998)
  [arXiv:hep-ph/9708209].

\bibitem{Barenboim:2001fd}
  G.~Barenboim, F.~J.~Botella and O.~Vives,
  Nucl.\ Phys.\  B {\bf 613}, 285 (2001)
  [arXiv:hep-ph/0105306].


\bibitem{Chen:2008ug}
  C.~H.~Chen, C.~Q.~Geng and L.~Li,
  Phys.\ Lett.\  B {\bf 670}, 374 (2009)
  [arXiv:0808.0127 [hep-ph]].

\bibitem{Mohanta:2008fa}
  R.~Mohanta and A.~K.~Giri,
  Phys.\ Rev.\  D {\bf 78}, 116002 (2008)
  [arXiv:0812.1077 [hep-ph]].


\bibitem{Langacker:2000ju}
  P.~Langacker and M.~Plumacher,
  Phys.\ Rev.\  D {\bf 62}, 013006 (2000)
  [arXiv:hep-ph/0001204].

\bibitem{Barger:2003hg}
  V.~Barger, C.~W.~Chiang, P.~Langacker and H.~S.~Lee,
  Phys.\ Lett.\  B {\bf 580}, 186 (2004)
  [arXiv:hep-ph/0310073];
\bibitem{Chiang:2006we}
  C.~W.~Chiang, N.~G.~Deshpande and J.~Jiang,
  JHEP {\bf 0608}, 075 (2006)
  [arXiv:hep-ph/0606122].

\bibitem{Barger:2009eq}
  V.~Barger, L.~Everett, J.~Jiang, P.~Langacker, T.~Liu and C.~Wagner,
  Phys.\ Rev.\  D {\bf 80}, 055008 (2009)
  [arXiv:0902.4507 [hep-ph]];

\bibitem{ZprPheno3}
  V.~Barger, L.~L.~Everett, J.~Jiang, P.~Langacker, T.~Liu and C.~E.~M.~Wagner,
  arXiv:0906.3745 [hep-ph].

\bibitem{ZprReview}
  P.~Langacker,
  Rev.\ Mod.\ Phys.\  {\bf 81}, 1199 (2009)
  [arXiv:0801.1345 [hep-ph]].

%




\bibitem{Yang:2010qd}
  K.~C.~Yang,
  arXiv:1010.2944 [hep-ph].


\bibitem{Cheng:2010hn}
  H.~Y.~Cheng, Y.~Koike and K.~C.~Yang,
  Phys.\ Rev.\  D {\bf 82}, 054019 (2010)
  [arXiv:1007.3541 [hep-ph]].



\bibitem{Keum:2000ph}
  Y.~Y.~Keum, H.~n.~Li and A.~I.~Sanda,
  Phys.\ Lett.\  B {\bf 504}, 6 (2001)
  [arXiv:hep-ph/0004004];
  Phys.\ Rev.\  D {\bf 63}, 054008 (2001)
  [arXiv:hep-ph/0004173];
  C.~D.~Lu, K.~Ukai and M.~Z.~Yang,
  Phys.\ Rev.\  D {\bf 63}, 074009 (2001)
  [arXiv:hep-ph/0004213];
  C.~D.~Lu, and M.~Z.~Yang,
  Eur.\ Phys.\ J.\  C {\bf 23}, 275-287 (2002)
  [arXiv:hep-ph/0011238].



\bibitem{Wang:2010tz}
  Z.~G.~Wang,
  arXiv:1011.3200 [hep-ph].





\bibitem{Chen:2010aq}
  C.~H.~Chen, C.~Q.~Geng and W.~Wang,
  JHEP {\bf 1011}, 089 (2010)
  [arXiv:1006.5216 [hep-ph]].











\bibitem{Buchalla:1995vs}
  G.~Buchalla, A.~J.~Buras and M.~E.~Lautenbacher,
  Rev.\ Mod.\ Phys.\  {\bf 68}, 1125 (1996)
  [arXiv:hep-ph/9512380].




\bibitem{HFAG}
  E.~Barberio {\it et al.}  [Heavy Flavor Averaging Group (HFAG)],
  arXiv:1010.1589. The updated results can be found at
  www.slact.stanford.edu/xorg/hfag.
%


\bibitem{Amsler:2008zz}
  K.~Nakamura  {\it et al.}  [Particle Data Group],
  J.\ Phys.\  G {\bf 37}, 075021 (2010).





\bibitem{Abazov:2010fs}
  V.~M.~Abazov {\it et al.}  [D0 Collaboration],
  Phys.\ Lett.\  B {\bf 693}, 539 (2010)
  [arXiv:1006.3469 [hep-ex]].





\bibitem{Buras:1994dj}
  A.~J.~Buras and M.~Munz,
  Phys.\ Rev.\  D {\bf 52}, 186 (1995)
  [arXiv:hep-ph/9501281].

\bibitem{Lim:1988yu}
  C.~S.~Lim, T.~Morozumi and A.~I.~Sanda,
  Phys.\ Lett.\  B {\bf 218} (1989) 343.

\bibitem{Ali:1991is}
  A.~Ali, T.~Mannel and T.~Morozumi,
  Phys.\ Lett.\  B {\bf 273}, 505 (1991).


\bibitem{Lu_and_zhang}
   C.~D.~Lu and D.~X.~Zhang,
  Phys.\ Lett.\  B {\bf 397}, 279 (1997)
  [arXiv:hep-ph/9702358].


\end{thebibliography}
\end{document}